\documentclass[10pt]{article}
\usepackage{amsmath, amsfonts,latexsym,color,array}
\newcommand{\R}{\mathbb R}

\newcommand{\N}{\mathbb N}
\newcommand{\C}{\mathbb C}

\newcommand{\calE}{{\mathcal E}}
\newcommand{\calC}{\mathcal C}
\newcommand{\calN}{\mathcal N}
\newcommand{\calM}{\mathcal M}

\newcommand{\calL}{\mathcal L}
\newcommand{\calG}{\mathcal G}

\newcommand{\calQ}{\mathcal Q}
\newcommand{\calT}{\mathcal T}
\newcommand{\calV}{\mathcal V}

\newcommand{\del}{\partial}

\newcommand{\calD}{{\mathcal D}}

\renewcommand{\div}{\mathop{\rm div}}
\newcommand{\dive}{{\mbox{\rm div\,}}}
\newcommand{\divo}[1]{{\mbox{\rm div}_{#1}\,}}
\newcommand{\tr}{{\mbox{\rm tr\,}}}
\newcommand{\curl}{\mathop{\rm curl}}

\newcommand{\Ord}[1]{{\mathcal O}\left(#1\right)}

\newcommand{\tgamma}{{\tilde{\gamma}}}    % Metric
\newcommand{\extd}{d\;}
\newcommand{\cf}{{\psi}}              % Conformal factor
\newcommand{\cfT}{{\psi_T}}

\newcommand{\mf}{{\mathcal F}}        % Matter field

\newcommand{\tmf}{{\tilde{\mf}}}

\newcommand{\mCnstr}{{\mathcal C}}    % Additional matter field constraints
\newcommand{\mBundle}{E}              % Bundle where matter fields take 
\newcommand{\mConf}{\Phi}             % Matter field conformal transformation
\newcommand{\laplacian}{\Delta\,}              % Laplacian operator
\newcommand{\laplaciano}[1]{\Delta_{#1}\,}     % Laplacian operator
\newcommand{\Reals}{\mathbb R}
\newcommand{\Ball}{\mathbb B}
\newcommand{\wnorm}[1]{\left|\left|#1\right|\right|_{k,\alpha,\delta}}
\newcommand{\norm}[1]{\left|\left|#1\right|\right|}
\newcommand{\abs}[1]{\left|#1\right|}
\newcommand{\mnl}{n}                   % Notation for ``matter induced 
\newcommand{\ep}{\wedge}
\newcommand{\ad}{\mathop{\rm ad}}
\newcommand{\dAd}[1][]{\mathop{{\rm D}^*_{#1}}}
\newcommand{\dop}[1][]{\mathop{{\rm D}_{#1}}}
\newcommand{\hs}[1][]{\mathbin{\ast_{#1}}}
\newcommand{\lieg}{\mathfrak{g}}
\newcommand{\fip}[2]{\left<#1,#2\right>}
\newcommand{\eop}{\hfill$\Box$}
\newcommand{\Vol}{{\mathop{\rm Vol}}}

\newcounter{shownewstuffflag}

\newcommand{\startnewstuff}{\ifnum\value{shownewstuffflag}>0\color{blue}\fi}
\newcommand{\finishnewstuff}{\ifnum\value{shownewstuffflag}>0\color{black}\fi}
\newcounter{oldeq}
\newcommand{\starteqgroup}[1]{\setcounter{oldeq}{\value{equation}}
\setcounter{equation}{0}\renewcommand{\theequation}{#1}
}
\newcommand{\finisheqgroup}{\renewcommand{\theequation}{\arabic{equation}}\setcounter{equation}{\value{oldeq}}}

\def\beq{\begin{equation}}
\def\eeq{\end{equation}}

\def\ip<#1,#2>{\left<#1,#2\right>}

\newtheorem{theorem}{Theorem}
\newtheorem{lemma}{Lemma}
\newtheorem{proposition}{Proposition}

\newtheorem{definition}{Definition}

\begin{document}
\title{A gluing construction for non-vacuum solutions of the Einstein 
constraint equations}

\author{James Isenberg\thanks{Partially supported by the NSF under Grants 
PHY-0099373, PHY-0354659 and the American Institute of Mathematics}
\\ University of Oregon  \and
David Maxwell\thanks{Partially supported by the NSF under Grant DMS-0305048
        and the UW Royalty Research Fund}
\\ University of Alaska Fairbanks \and
Daniel Pollack\thanks{Partially supported by the NSF under Grant DMS-0305048
        and the UW Royalty Research Fund}
\\ University of Washington}

\date{\today}
%\date{Almost Final Draft: DP January 20, 2005}

\maketitle

\begin{abstract}
We extend the conformal gluing  construction of \cite{IMP01} 
by establishing an analogous gluing result for field theories obtained by
minimally coupling Einstein's gravitational theory with matter fields.
We treat classical fields such as perfect fluids and the Yang-Mills 
equations as well as the Einstein-Vlasov system, which is an important example
coming from kinetic theory.  In carrying out these extensions, we extend
the conformal gluing technique to higher dimensions and 
codify it in such a way as to make more transparent where it can, 
and can not, be applied. In particular, we show exactly 
what criteria need to be met in order to apply the construction, in
its present form, to any 
other non-vacuum field theory.
\end{abstract}

\section{Introduction}
\label{se:intro}

One of the effective tools which have been recently developed 
and employed for the 
construction and study of initial data for solutions of Einstein's 
gravitational field equations is the method of gluing.
This is a technique which has had a long and fruitful history in geometric
analysis but which has only recently been 
successfully applied to general relativity.
The idea of the gluing method (in its simplest form) is that, given a pair 
of sets of initial data which 
satisfy the Einstein constraint equations, we may use it to construct 
a new set of initial data which a) lives on a connected sum of the 
manifolds of the given sets of data, b) solves the constraint  equations, 
and c) closely approximates the original sets of data on the parts of the 
new manifold which correspond to the original manifolds 
(i.e., away from the  tubular ``neck" of the connected sum).  
The work of Isenberg, Mazzeo, and Pollack \cite{IMP01}, \cite{IMP03}, 
shows that this sort of  
gluing can be carried out for fairly general sets of {\it vacuum} 
initial data. That work also describes a number of applications 
of gluing, such as producing multi-black hole initial data, 
adding wormholes to given sets of data, and showing that an 
arbitrary closed manifold with a point removed always admits 
both asymptotically Euclidean and asymptotically hyperbolic 
solutions of the vacuum constraint equations.

In this work, we show that the gluing results and the gluing 
applications which are discussed  for vacuum initial data in 
\cite{IMP01}  can be extended to  fairly general 
sets of non-vacuum data as well. We do this here for a number of 
special cases, including Einstein-Maxwell, 
Einstein-Yang-Mills, Einstein-fluids, and Einstein-Vlasov, as well as
for any of these theories with a cosmological constant added.
We also discuss the features which a general field theory 
should have if, when it is coupled to Einstein's equations, 
solutions of the corresponding constraint equations should allow gluing. 

As for vacuum data, the gluing procedure applied to non-vacuum data 
relies quite  heavily on the conformal method for obtaining solutions 
of the constraint equations. Thus, after commenting in Section 2 on 
the general form of the constraint equations for non-vacuum field 
theories, we proceed in that section to describe the application 
of the conformal method to such theories. For the non-vacuum theories 
listed above--Einstein-Maxwell, etc--the conformal method leads to 
determined sets of (nonlinear) elliptic equations which, at least 
for constant mean curvature (``CMC") data, are readily analyzed for 
solubility. This is not true for all non-vacuum field theories; 
indeed, for any non-vacuum field theory which involves derivative 
coupling (e.g., the Einstein-vector-Klein-Gordon theory), the 
conformal method leads to equations which are intractable using 
known techniques. (See \cite{IN77}.) Thus the gluing procedure 
which we use does not work for such theories. 

Note that in our discussion of the conformal method, we focus 
on the situation in which the initial data has constant mean 
curvature. We do this because, as with the vacuum case, even when  
the constraint solutions we are gluing together have non-constant 
mean curvature, the analysis we rely on to carry out the gluing 
is based primarily on the CMC version of the conformal treatment 
of the constraints.  (See \cite{IMP03}.)

While the details of the analysis differ from one non-vacuum field 
theory to another, the basic steps of the gluing procedure are 
largely the same for those non-vacuum field theories with the 
appropriate form for the constraint equations (in conformal form).  
Thus for these field theories,  we can  present a general 
discussion of these basic steps: (i) conformal blowup at the 
gluing points; (ii) connected sum of the manifolds and patching 
of the conformal metrics; (iii) patching of the non-gravitational 
fields,  solution of the non-gravitational constraints, and 
deformation estimates; (iv) patching of the extrinsic curvatures, 
solution of the momentum constraint and deformation estimates; 
(v) patching of the conformal factor, solution of the Hamiltonian 
constraint (in Lichnerowicz form) and deformation estimates; 
(vi) conformal recomposition of the initial data, forming the 
glued solution. We do this in Section 3. Also in that section, 
we outline the general analysis which leads to a proof that the 
gluing can be carried out  for appropriate sets of initial data 
for these theories. The section culminates with a general 
statement of our gluing results for general theories.

We discuss some of the details of gluing for various particular 
non-vacuum field theories in Section 4. Included are discussions 
of the field theories listed above.  However, to avoid repetition, 
we focus primarily on two of them: Einstein-Perfect-Fluids and 
Einstein-Yang-Mills.

Although the physically important versions of most of the 
field theories we discuss here are defined on 3+1 dimensional spacetimes,  
our results hold for arbitrary dimension. Hence we state most  of our 
formulations and our results for $n+1$ dimensional spacetimes where $n\ge 3$.

A very different and important type of gluing construction has been
developed by Corvino and Schoen \cite{C00}, \cite{CS03} 
and adapted and applied by Chru\'sciel and Delay \cite{CD02}, \cite{CD03}.  
These results exploit the 
underdetermined nature of the constraint equations as opposed to using
the conformal method to convert them into a determined system. This has
lead to a number of remarkable results beginning with the existence of 
a large class of asymptotically Euclidean spacetimes which are exactly 
Schwarzchild near infinity. Recently, by combining these techniques with 
the results of \cite{IMP01} and with the previous work of Bartnik \cite{B88}, 
Chru\'sciel, Isenberg and Pollack \cite{CIP-PRL}, \cite{CIP04}
have obtained  a gluing construction of the type 
described here which is  optimal in two distinct ways. First,
it applies to {\it generic} initial data sets and the required
(generically satisfied) hypotheses are geometrically and physically natural.
Second, the construction is completely {\it local} in the sense that the
initial data is left
unaltered on the complement of arbitrarily small neighborhoods of the
points about which the gluing takes place.  Using this construction they
have been able to establish the existence of cosmological, 
maximal globally hyperbolic, vacuum
space-times with no constant mean curvature spacelike Cauchy surfaces.
 Except for the case of generic non-gravitational 
fields described entirely by an energy density function $\rho$ and 
a current density vector field $J$ (satisfying a strict energy 
condition $\rho > |J|$), the Corvino-Schoen techniques have not 
yet been generalized away from the vacuum case. It is, however, 
expected that this can be done; it would then follow that the 
gluing theorems obtained for vacuum data in \cite{CIP04} would 
extend to non-vacuum data.

We end this introduction by remarking that we have, for simplicity, 
restricted ourselves here
to the consideration of initial data on compact manifolds (i.e., the 
{\em cosmological} setting).  The gluing results presented here have analogous 
statements which are valid for either asymptotically Euclidean or 
asymptotically hyperbolic initial data sets.  The required adaptations,
which are similar to those discussed in detail in \cite{IMP01}, are left
to the interested reader.

\section{The Constraint Equations and the Conformal Method}
\label{se:confmethod}

We restrict our attention in this work to classical field theories which 
are obtained by minimally coupling a spacetime covariant field theory to 
Einstein's gravitational theory, and which have a well-posed Cauchy 
formulation.  For such theories, if one is given a set of initial data 
which satisfies the constraint equations corresponding to that theory, 
one can always evolve to obtain a spacetime solution of the full PDE system. 
Our main interest here is primarily in the construction of solutions of 
the constraints.

For the theories we are interested in here, the initial data consist 
of a choice of  an $n$-dimensional manifold $\Sigma^n$ together with a 
Riemannian metric $\gamma$, a symmetric tensor $K$, and a collection of 
non gravitational fields which we collectively label $ \mf$, all 
specified on $\Sigma^n$. These non gravitational fields are usually, 
but not always, sections of a bundle over $\Sigma^n$.  The constraint 
equations which these initial data must satisfy generally take the form 
\begin{eqnarray}
\dive K - \extd \tr K & = & J(\mf,\gamma)  \label{eq:c1}\\
R_\gamma - |K|^2_\gamma + (\tr K)^2 & = & 2\rho(\mf,\gamma) \label{eq:c2}\\
\mCnstr(\mf, \gamma) & = & 0, \label{eq:c3}
\end{eqnarray}
where $J$ is the current density of the non-gravitational fields, 
$\rho$ is their energy density,
and $\mCnstr$ denotes  the set of additional constraints that come 
from the non gravitational part of the theory.\footnote{If we use $T$ 
to denote the stress-energy tensor for the non gravitational fields 
and we use $e_{\perp}$ to denote the unit normal to the hypersurface 
$\Sigma$ embedded in the spacetime solution generated from the initial 
data $(\Sigma^n, \gamma, K, \mf)$, then $J= - T(e_\perp,\,)$  and 
$\rho=T(e_\perp, e_\perp)$. Note that we have chosen units so that 
$8\pi G=1=c$.} Note that the first of these constraint equations is 
known as the momentum constraint,  the second is often referred 
to as the Hamiltonian constraint, while the last are collectively 
labeled the non-gravitational constraints.

As an example, for the Einstein-Maxwell theory in 3+1 dimensions, 
the non-gravitational fields consist of the electric and magnetic 
vector fields $E$ and $B$, we have 
$\rho=\frac{1}{2}(|E|^2_\gamma +|B|^2_\gamma$), $J=(E\times B)_\gamma$, 
and we have the extra (non-gravitational) constraints 
$\dive_{\gamma} E=0$ and $\dive_{\gamma} B=0$.

The system of constraint equations (\ref{eq:c1})-(\ref{eq:c3}), 
vacuum or non-vacuum, is an underdetermined PDE system. The idea 
of the conformal method is to split the initial data fields into 
two sets of fields: the ``conformal data", which is freely chosen, 
and  the ``determined data"  which is to be found by solving the  
constraints. In the familiar vacuum case \cite{CBY80},  the 
conformal data consists of the manifold $\Sigma^n$, a  Riemannian 
metric $\gamma$, a divergence-free trace-free symmetric tensor 
$\sigma$, and a function $\tau$, while the determined data consists 
of a positive definite function $\phi$ and a vector field $W$. 
With the conformal data  $(\Sigma^n, \gamma, \sigma, \tau)$ chosen, 
one determines $(\phi, W)$ by solving the equations
\beq
\divo{\gamma}(\calD W)=\frac{n-1}{n}\phi^{q+2} \nabla \tau
\label{vacmom:noncmc}
\eeq
and 
\beq
\label{vacham:noncmc}
\laplaciano{\gamma} \phi - \frac{1}{q(n-1)} R_\gamma\phi +
\frac{1}{q(n-1)}\left|\sigma + \calD W \right|^2_\gamma 
\phi^{-q-3}-\frac{1}{qn}\tau^2\cf^{q+1}=0,
\eeq
where $\calD W$ is the conformal Killing operator, with coordinate
representation 
\beq
\calD W_{ab} = \nabla_a W_b + \nabla_b W_a - \frac{2}{n}\gamma_{ab} 
\nabla_c W^c,
\label{LWeq}
\eeq
and where $q=\frac{4}{n-2}$ is a dimensional constant. If for a 
given choice of the conformal data one does determine $W$ and $\phi$
satisfying (\ref{vacmom:noncmc})-(\ref{vacham:noncmc}), 
then the fields 
\begin{eqnarray}
\tilde \gamma & = & \phi^q \gamma \label{gammaeq}\\
\tilde K & = & \phi^{-2} (\sigma + \calD W) + \frac{\tau}{n} 
 \phi^q \gamma  \label{Keq}
\end{eqnarray}
satisfy the vacuum constraints, consisting of (\ref{eq:c1})-(\ref{eq:c2}) 
with vanishing $\rho$ and $J$.
Note that the explicit form of the equations (\ref{vacmom:noncmc}) 
and (\ref{vacham:noncmc}) is determined by the form of the field 
decomposition, expressed in (\ref{gammaeq})-(\ref{Keq}). The explicit 
choice of the form of the field decomposition is in turn determined 
to a large extent by the two identities (for $\tilde \gamma=\phi^q \gamma$)
\beq
R_{\tilde \gamma} = -\phi^{-q-1} 
(q(n-1)\Delta_\gamma \phi - R_{\gamma}\phi) 
\label{ConfReq}
\eeq
(with $q=\frac{4}{n-2}$ being the unique exponent which avoids 
$|\nabla \phi|^2$ terms in (\ref{ConfReq}))
and 
\beq
\nabla_{\tilde \gamma}^a (\phi^{-2} B_{ab}) = \phi^{-q-2} 
\nabla _{\gamma}^a B_{ab}
\label{confdiveq}
\eeq 
which holds for any trace-free tensor $B$.

The extent to which there exist unique solutions to equations 
(\ref{vacmom:noncmc}) and (\ref{vacham:noncmc}) for various classes 
of conformal data has been studied extensively; see \cite{BI04} for 
a recent review. Here, we note primarily that while the issue is 
fairly well understand for constant mean curvature (``CMC") conformal 
data and near-CMC conformal data, very little is known more generally. 
Consequently, the earliest gluing results for the vacuum constraints 
\cite{IMP01} pertain to CMC data sets, and later results rely 
primarily on CMC analysis. Hence we focus on CMC data sets here. 

A set of initial data has constant mean curvature if 
$\tr\tilde K = \tau$ is constant on $\Sigma^n$. This condition 
significantly simplifies the analysis of the vacuum constraint 
equations because it decouples equations  (\ref{vacmom:noncmc}) 
and (\ref{vacham:noncmc}). Since $W\equiv 0$ is a solution 
to (\ref{vacmom:noncmc}) with vanishing right hand side, the analysis reduces 
to first finding a divergence-free trace-free symmetric tensor, and 
then solving (\ref{vacham:noncmc}) (often referred to as the 
``Lichnerowicz equation'') with $\calD W=0$.

To extend the conformal method to the non-vacuum constraints, 
with non-gravitational fields $\mf$ present, one needs to extend 
the field decomposition (\ref{gammaeq})-(\ref{Keq}) to $\mf$. The 
chief criteria generally used to decide how to do this are the 
following \cite{IN77}: (I) In the CMC case, the constraint system 
should be semi-decoupled, in the sense that one can first solve the 
non-gravitational constraints independently of $W$ and $\phi$, and 
then one can solve the momentum constraint for $W$ independently 
of $\phi$, and then finally one solves the Lichnerowicz equation 
for $\phi$. (II) The addition of the non-gravitational terms to 
the Hamiltonian constraint should not result in the Lichnerowicz 
equation containing either derivatives 
of $\phi$ or those powers of $\phi$ which would  lead to 
insurmountable difficulties in the subsequent analysis.

To make these criteria more precise, let us presume that the field 
decomposition for the non-gravitational fields is defined by an 
action $\Phi$ of the group of conformal factors ($C^\infty_+(\Sigma)$ 
under multiplication) on the set of matter fields, so that the physical 
fields $\tmf$, which together with $ \tgamma$ and $\tilde K$ must satisfy 
the constraints (\ref{eq:c1})-(\ref{eq:c3}), are given by 
$\tmf = \Phi(\mf, \phi)$ where $\mConf(\mf,1)=\mf$ and 
$\mConf(\mConf(\mf,\phi_1),\phi_2)=\mConf(\mf,\phi_1\phi_2)$. 
Note that the data $\mf$ is included in the set of conformal data 
(along with $\gamma, \sigma$, and $\tau$); the explicit form of the 
action $\Phi$ is to be chosen for each theory. 

In this language, the first of our criteria is satisfied so long as 
$\mCnstr, \Phi$, and $J$ satisfy the conditions

\starteqgroup{C\arabic{equation}}
\beq
\mCnstr\left(\mConf(\mf,\phi),\phi^q \gamma  \right)= 
\phi^p \mCnstr\left(\mf,\gamma \right) 
\label{eq:Conf1}
\eeq
and
 \beq
 J(\mConf(\mf,\phi),\phi^q\gamma) = \phi^{-q-2} J(\mf,\gamma),
\label{eq:Conf2}
\eeq
\finisheqgroup for some number $p$. 
As a consequence of (\ref{eq:Conf1}),  if we choose the conformal data 
$( \gamma, \sigma, \tau, \mf)$ so that $\mCnstr\left(\mf,\gamma\right)=0$, 
then whatever $\phi$ and $W$ are determined to be, the constraint  
$\mCnstr\left(\tmf,\tgamma\right)=0$ is satisfied by the physical 
initial data. As a consequence of (\ref{eq:Conf2}), the conformal form 
of the momentum constraint is (compare (\ref{vacmom:noncmc}))
\beq
\divo{\gamma}(\calD W)=\frac{n-1}{n}\phi^{q+2} \nabla \tau + J(\mf, \gamma),
\label{nonvacmom:noncmc}
\eeq
which in the CMC case can be solved for $W$ independent of $\phi$.

Satisfaction of the second criteria depends upon the form of the term 
$\rho (\mConf(\mf,\phi), \phi^q\gamma)$. It is crucial first of all 
that this quantity involves neither $W$ nor any derivatives of $\phi$. 
We formalize this by assuming that, for each choice of the matter field  
$\mf$ and the metric $\gamma$, there exists a function  
$\mnl_{\mf,\gamma}:\Reals^+\times\Sigma\rightarrow\Reals$
such that at any point $p$ of $\Sigma$,
\starteqgroup{{C\arabic{equation}}}
\setcounter{equation}{2}
\beq
\phi^{q+1}2\rho(\mConf(\mf,\phi),\phi^{q}\gamma) 
= n_{\mf,\gamma}(\phi(p),p).
\label{eq:Conf3}
\eeq  
\finisheqgroup
Note that in practice, to ensure that this condition holds for a given 
field theory, it is generally sufficient that the non-gravitational 
fields do not involve derivative coupling. 

Assuming that condition (\ref{eq:Conf3}) holds, the Lichnerowicz equation 
takes the form 
\beq
\label{nonvacham:noncmc}
\laplaciano{\gamma} \phi -a_1 R_\gamma\phi +
a_1\left|\sigma + \calD W \right|^2_\gamma \phi^{-q-3} + 
-a_2\tau^2\phi^{q+1}+ a_1 n_{\mf,\gamma}(\phi) =0, 
\eeq
where for convenience we set $a_1= \frac{1}{q(n-1)}=\frac{n-2}{4(n-1)}$ and  
$a_2=\frac{1}{qn}=\frac{n-2}{4n}$, and also for convenience we suppress 
the dependence of $n_{\mf,\gamma}$ on its second argument. To ensure that 
this equation is analytically tractable, it is important that for fixed 
$\gamma$ and $\mf$,  $n_{\mf,\gamma}(\phi) $  is a monotonically 
decreasing function of $\phi$. In practice, we find that this condition 
is often satisfied  by $n_{\mf,\gamma}(\phi)$ being expressible as a 
sum of negative powers of $\phi$, with non-negative coefficients.

While the two criteria and their consequent conditions 
(\ref{eq:Conf1})-(\ref{eq:Conf3})  on $\mConf(\mf,\phi)$ 
and its interaction with 
$\mCnstr, J$, and $\rho$ appear quite restrictive, it is shown in 
\cite{IN77} that for most familiar physical fields, a choice of 
$\mConf(\mf,\phi)$ which satisfies these criteria can be found. 
We shall discuss a number of such examples in this paper. 
Before proceeding, we illustrate how this works for a simple 
example: Einstein-Maxwell in three  space dimensions .

The conformal data for the Einstein-Maxwell theory consists of 
$(\Sigma^3, \gamma, \sigma, \tau, B, E)$, where  
$(\Sigma^n, \gamma, \sigma, \tau)$ are  the usual vacuum conformal 
data, and both $B$ and $E$ are vector fields which are required to 
be divergence-free with respect to the metric $\gamma$.  
To satisfy the two criteria, we choose  $\mConf(B^a,\phi)=B^a \phi^{-6}$ 
and $\mConf(E^a,\phi)=E^a\phi^{-6}$. It follows then that (i) the 
extra constraints $\dive_{\gamma} \tilde B=0$ and $\dive_{\gamma} 
\tilde E=0$ are satisfied automatically  so long as $B$ and $E$ 
are both divergence-free with respect to the metric $\gamma$; 
(ii)  the momentum constraint takes the form 
\beq
\divo{\gamma}(\calD W)=\frac{2}{3}\phi^6 \nabla \tau + (E \times B),
\label{EinMaxmom:noncmc}
\eeq
which is independent of $\phi$ in the CMC case; and (iii) the Lichnerowicz 
equation takes the form 
\beq
\label{EinMaxham:noncmc}
\laplaciano{\lambda} \phi - \frac{1}{8} R_\gamma\phi +
\frac{1}{8}\left|\sigma + \calD W \right|^2_\gamma \phi^{-7} 
- \frac{1}{12}\tau^2\phi^{5}+ \frac{1}{8} (E^2 + B^2)  \phi^{-3} =0.
\eeq
We note that the extra term in (\ref{EinMaxham:noncmc}) involves $\phi$ 
with a negative power and a positive coefficient, much like the 
$|\sigma+LW|$ term. As discussed in \cite{I95}, existence and uniqueness 
of solutions of (\ref{EinMaxmom:noncmc})-(\ref {EinMaxham:noncmc}) 
in the CMC case then closely follows the pattern of the vacuum case. 

\section{The Gluing Construction for Non-Vacuum Solutions}
\label{se:GlueCon}
\subsection{Overview}
\label{sse:Over}
Before generalizing the gluing construction from \cite{IMP01} to apply to 
non-vacuum fields, we
briefly summarize the original technique, modified to arbitrary spatial 
dimension $n \geq 3$.  We start with
an $n$-manifold $\Sigma$ and a CMC solution $(\gamma,K)$ of the
vacuum Einstein constraint equations (so $K=\sigma+\frac{\tau}{n}\gamma$, 
with $\sigma$ divergence-free and trace-free, and with $\tau$ constant).
We fix two points $p_1$ and $p_2$ of $\Sigma$ and a small radius $R$. 
Let $B_j=B_R(p_j)$ be the balls on which we will do surgery, let
$\Sigma^*=\Sigma\backslash\{p_1,p_2\}$ and let  
$\Sigma^*_r = \Sigma\backslash(B_r(p_1)\cup B_r(p_2))$.  The 
construction then proceeds as follows.
\begin{itemize}
\item We first construct a conformally related metric on $\Sigma^*$ 
agreeing with $\gamma$ away from the surgery site and having
two asymptotically cylindrical ends at the puncture locations.  
Let $\cf_c$ be a conformal factor equal to 1 on $\Sigma^{*}_{2R}$ and 
equal to 
$r_j^{2/q}$ on $B_j$, where $r_j$ is the geodesic distance from $p_j$, 
and where $q=\frac{4}{n-2}$ as before. Then 
$\gamma_c = \cf_c^{-q}\gamma$ is the desired metric. Setting 
$\sigma_c=\cf_c^{2}\sigma$ and $K_c=\sigma_c+\frac{\tau}{n}\gamma$
we see that $(\gamma_c,K_c)$ satisfies the momentum constraint
and that $\cf_c$ satisfies the Lichnerowicz equation with respect to 
$(\gamma_c,K_c)$.
\item We next perform surgery on the cylindrical ends by identifying
finite segments of length $T$ to construct a family of topologically 
identical manifolds $\Sigma_T$. To do this, we first construct maps 
from $B_j\backslash p_j$ to 
the half cylinder $(0,\infty)\times S^{n-1}$ by sending points at the ball
radius $r_j$ to the cylinder length $t_j = -\log r_j + \log R$ and by using 
Riemann normal  coordinates to determine the projections onto $S^{n-1}$. 
We then identify the finite segments $\{(t_j,\theta):0<t_j<T\}$ 
via the map $(t_1,\theta)\mapsto(T-t_1,-\theta)$, resulting in the
smooth manifold $\Sigma_T$.  Letting $s=t_1-T/2=T/2-t_2$,
we denote by $Q_{l,a}$ the cylindrical segment
$\{(s,\theta):a-l<s<a+l\}$ of length $2l$ centered at $a$. We use the 
shorthand notation $Q_{l}=Q_{l,0}$ for centered segments, $Q=Q_{1}$ for a 
short collar in the middle, and $C_{T}=Q_{T/2}$ for the entire 
cylindrical region.
\item We now construct approximate solutions $(\gamma_T,K_T,\cf_T)$ of 
the momentum constraint and the Lichnerowicz equation on $\Sigma_T$ 
by using 
the conformally modified solution away from the surgery site and using cutoff 
functions to piece together an approximation along the identified 
cylindrical segment. 
\item In preparation for using the CMC-conformal technique, to map 
the approximate solutions to full solutions, 
we perturb the trace free part of $K_T$ to obtain a constant
trace second fundamental form 
$\hat K_T=\hat \sigma_T+\frac{\tau}{n}\gamma_T$ which, 
together with $\gamma_T$, satisfies the momentum constraint.  
\item Finally, we solve the Lichnerowicz equation using a contraction-map
argument to arrive at a $T$ parameterized set of solutions of the 
constraints which, for large $T$,
is ``close'' to the original one away from the surgery site.  
\end{itemize}

The last step is quite delicate; the contraction-map argument 
demands some analytic conditions on the linearization
$\calL_T$ about $\cf_T$ of the Lichnerowicz operator $\calN_T$ on $\Sigma_T$.
In particular, $\calL_T$ must be surjective, and there must
exist bounds uniform in $T$ for the norm of its inverse on 
certain weighted H\"older spaces.  The argument also imposes 
some stringent conditions on the size of the error terms which arise 
from the earlier
approximations and corrections.  Substantial work in
\cite{IMP01} is devoted to constructing and obtaining precise estimates for
the perturbation of $\sigma_T$ to $\hat \sigma_T$ that enter into this analysis.

To extend the technique to include matter fields, we start as before with
a CMC solution $(\gamma,K,\mf)$ of the Einstein-matter constraints
on an $n$-manifold $\Sigma$ and a fibre bundle $\mBundle$ over 
$\Sigma$.
We also assume that we have chosen a conformal group action $\mConf$ for 
the non gravitational  fields which satisfies the criteria discussed 
in Section \ref{se:confmethod}.

The topological step that constructs $\Sigma_T$ must generally be 
supplemented with a construction of appropriate fibre bundles 
$\mBundle_T$ over $\Sigma_T$.  Since each of the balls $B_i$ is 
contractible, there exists a local trivialization
over each of them.  We can then use these trivializations to identify
fibres: if $q_1\in B_1$ is identified with $q_2\in B_2$ in the connected 
sum to create $\Sigma_T$, we can identify the fibre over $q_1$ with the 
fibre over $q_2$ via our pair of fixed local trivializations.
On the other hand, if $q\in\Sigma_T$ is outside of the neck, then we take
the fibre to be the one over $q$ in $\mBundle$. The result is a smooth fibre 
bundle $\mBundle_T$ over $\Sigma_T$.

Together with the conformally modified gravitational fields $\gamma_c$ 
and $K_c$,
we define on $\Sigma^*$ the conformally modified non-gravitational fields 
\[
\mf_c=\mConf(\mf,\cf_c^{-1}).
\]
Then as a consequence of the choice of $\mConf$, we verify that 
$(\gamma_c,K_c,\mf_c)$ satisfy the non-gravitational and momentum 
constraints on $\Sigma^*$, and in addition $\psi_c$ satisfies the 
Lichnerowicz equation corresponding to this data:
\beq
\label{Lichnero_c}
\laplaciano{\gamma_c}  \psi_c -a_1 R_{\gamma_c} \psi_c + 
a_1 | \sigma_c + \calD W |^2_{\gamma_c} \psi_c^{-q-3} 
-a_2\tau^2\psi_c^{q+1}
+ a_1 n_{\mf_c,\gamma_c}(\psi_c) =0, 
\eeq

To construct from $(\gamma_c,K_c,\mf_c)$ on $\Sigma^*$ a parameterized 
set of conformal data $(\gamma_T,K_T,\mf_T)$ on $\Sigma_T$, we use a 
cut-off function procedure as in the vacuum case. That is, we first 
set $(\gamma_T,K_T,\mf_T)$ = $(\gamma_c,K_c,\mf_c)$ on 
$\Sigma_T\backslash Q$. Then on $Q$ (recalling the definition of 
$s$ in terms of $T$) we let $\chi(s)$ be a cut-off function on $\R$ 
equal to 0 for $s>1$ and equal to 1 for $s<-1$, and we define 
(i) $\gamma_T=\chi(s)\gamma_1+(1-\chi(s))\gamma_2$; 
(ii) $\sigma_T=\chi(s)\sigma_1+(1-\chi(s))\sigma_2$ in $Q$ and 
$K_T=\sigma_T+\frac{\tau}{n}\gamma_T$ (recall that $\tau$ is a 
constant); and finally (iii) $\mf_T=\chi(s)\mf_1+(1-\chi(s))\mf_2$.  
Note that $\gamma_i, \sigma_i$, and $\mf_i$ (for $i \in \{1,2\}$) 
are all defined by identifying $C_T$ with a subset first of $B_1$ 
and then of $B_2$. If we do the same with $\psi_c$, and then set 
$\cf_T=\chi(t_2-1)\cf_1+\chi(t_1-1)\cf_2$ in $C_T$ and 1 outside, 
then we can verify (as discussed in further detail below) that  
for each value of $T$, $(\gamma_T,K_T,\mf_T)$ together with 
$\psi_T$ constitute an approximate solution of the constraints, 
including the Lichnerowicz equation. Note that 
$\psi_T=\psi_1+\psi_2$ on most of $C_T$.

To go from these approximate solutions to a parameterized set of 
exact solutions, we proceed as follows: 
First we perturb $\mf_T$, and thereby obtain a set of matter fields 
$\hat\mf_T$ which closely approximate 
$\mf_c$ outside the neck, and which together with $\gamma_T$ solve 
the non gravitational constraints globally on $\Sigma_T$.  
Then, we perturb $K_T$ in order to obtain a set of  
symmetric tensors $\hat K_T$ which closely approximate $K$ outside 
the neck and which together with  $\gamma_T$ and $\hat\mf_T$ 
satisfy the momentum constraint globally on $\Sigma_T$.
Note that both of these two perturbations involve solving partial 
differential equations for the perturbation terms (in order to satisfy the 
non gravitational constraints and the momentum constraint respectively).
The solvability of these equations (for the perturbation terms, given
the approximate solutions) is an issue that must be addressed at this stage.
Finally, provided that various 
error terms in the Lichnerowicz equation have been kept under
control after substituting in $\gamma_T$,  $\hat K_T$,  $\hat\mf_T$, 
and $\psi_T$, we use a contraction mapping argument to show that 
there exists a solution $\hat \psi_T$ to this equation with the 
conformal data ($\gamma_T$, $\hat K_T$, $\hat\mf_T$);  and we 
show further that away from the gluing region $Q$, the solution 
data ($\tilde \gamma_T=\hat \psi_T^q\gamma_T$, 
$\tilde K_T= \hat \psi_T^{-2}
\hat \sigma_T+\frac{1}{n}\hat \psi_T^q\gamma_T \tau$, 
$\tilde{\mf_T}=\Phi ( \hat\mf_T, \hat \psi_T)$) 
approaches the original data $(\gamma, K, \mf)$ arbitrarily closely. 

We discuss some of the details of these steps for generic non 
gravitational fields in the rest of this section, and discuss 
them for particular examples in section 4. 

\subsection{Satisfying the Non-Gravitational Constraints}

Since we presumably have chosen the action of the conformal map 
$\Phi$ on the matter fields so that the non-gravitational constraints 
decouple from the others in the conformal representation, the first 
step in going from $(\gamma_T,K_T,\mf_T)$ to ($\gamma_T$, $\hat K_T$, 
$\hat{\mf_T}$) is to choose $\hat{\mf_T}$ so that  
$\mCnstr(\hat{\mf_T},\gamma_T)=0$. In obtaining $\hat{\mf_T}$, 
we want it to be arbitrarily close (for $T$ sufficiently large)  
to $\mf_T$ on $\Sigma^*_R$, 
in order to minimize the errors that are introduced into the other 
constraints. In particular, it is crucial for the gluing procedure 
that the error estimates (\ref{eq:m1})-(\ref{eq:m2}), 
(\ref{eq:e1})-(\ref{eq:e2}), and  (\ref{eq:n1})-(\ref{eq:n4}) 
described below be satisfied.

The details of the construction of $\hat{\mf_T}$ are field 
specific; we discuss a number of cases in section 4. Here, we 
note what happens in the Einstein-Maxwell case: While the original 
$E$ and $B$ fields are divergence-free with respect to $\gamma$, 
and consequently the conformally mapped fields $E_c$ and $B_c$ 
are divergence-free with respect to $\gamma_c$, the fields $E_T$ 
and $B_T$ constructed using cut-off functions are not divergence-free  
with respect to $\gamma_T$ (or any metric). Effectively the 
non-gravitational constraints are equivalent to this divergence-free 
property. We obtain new fields $\hat E_T$ and $\hat B_T$ which satisfy 
the conditions $\divo{\gamma_T}\hat E_T=0$ and 
$\divo{\gamma_T}\hat B_T =0$ by carrying through the standard linear 
procedure. That is, we solve the linear equation
\beq
\label{nueq}
\Delta_{\gamma_T} \mu_T = \divo{\gamma_T} E_T
\eeq
for the scalar $\mu_T$, and then set 
\beq
\label{hatE}
\hat E_T=E_T-\nabla \mu_T;
\eeq
the divergence-free condition immediately follows. We carry out a 
similar procedure to obtain $\hat B_T$. Noting that the supports of 
$\divo{\gamma_T} E_T$ and of $\divo{\gamma_T} B_T$ are contained in 
$Q$, one can carry through the analysis which verifies the estimates 
(\ref{eq:m1})-(\ref{eq:m2}), (\ref{eq:e1})-(\ref{eq:e2}), and  
(\ref{eq:n1})-(\ref{eq:n4}) as detailed below.

\subsection{Repairing the Momentum Constraint}
\label{sse:momentum}

With $\hat \mf_T$ determined, we next need to find $\hat K_T$  
for which the momentum constraint is satisfied. We may do this 
by finding a symmetric trace-free (0,2) tensor $\hat \nu_T$ which 
satisfies 
\beq
\divo{\gamma_T} (\hat \nu_T) = J(\hat \mf_T,\gamma_T) - 
\divo{\gamma_T}\sigma_T.
\label{eq:mpert}
\eeq
If we can obtain a tensor $\hat \nu_T$ which satisfies this condition, 
and if we then set 
$\hat K_T = \sigma_T+\hat \nu_T+\frac{\tau}{n}\gamma_T$, then we do 
have a solution of the momentum constraint. 

To solve this equation for $\hat \nu_T$,  we let $\calD$ be the 
($\gamma_T$ compatible) conformal Killing operator
on vector fields $X$, so  
$\calD X = \calL_X\gamma_T-\frac{2}{n}\divo{\gamma_T}(X)\gamma_T$
where $\calL_X$ is the Lie derivative.  
Its formal adjoint $\calD^*$ is $-\divo{\gamma_T}$ and we set 
$L = \calD^*\calD$.  If $W_T$ is a vector field solving 
$$
L W_T = J(\hat \mf_T,\gamma_T) - \divo{\gamma_T}\sigma_T
$$ 
(where  we freely identify,
via $\gamma_T$, vectors and covectors), then we can set 
$\hat \nu_T=-\calD W_T$
to obtain a solution to (\ref{eq:mpert}), and consequently a solution 
$\hat K_T$ to the momentum constraint.  

Besides obtaining $\hat \nu_T$, we need to establish control of its 
size.  Lemma \ref{le:vlest} below, 
proved in \cite{IMP01} for $3$-manifolds $\Sigma$, provides this 
estimate in terms of the following H\"older norm.
\begin{definition}
\label{def:holder}
Let $||X||_{k,\alpha,\Omega}$ denote the H\"older norm (computed with 
respect to the metric $\gamma_T$) of the vector field $X$ on an open 
subset $\Omega$ of $\Sigma$.
Then we define
\[
||X||_{k,\alpha} = ||X||_{k,\alpha,\Sigma^*_{R/2}}+
\sup_{-\frac{T}{2}+1 \leq a \leq \frac{T}{2}-1} ||X||_{k,\alpha,Q_{1,a}}.
\]
\end{definition}

\begin{lemma} 
\label{lemma1}
Suppose there are no conformal Killing fields that 
vanish at the points $p_j$ of $\Sigma$.  Then 
for $T$ sufficiently large and for each
$X \in {\mathcal C}^{k,\alpha}(\Sigma_T)$ there is a unique
solution $W \in {\mathcal C}^{k+2,\alpha}(\Sigma_T)$ to
$L_{}W = X$.  Moreover, there exists a constant $C$ independent of 
$W$ and $T$ such that
\[
||W||_{k+2,\alpha} \leq C T^{3} ||X||_{k,\alpha}.
\]
\label{le:vlest}
\end{lemma} 
The proof of  Lemma \ref{lemma1} in general dimensions, 
which we skip now for the sake of exposition, is presented in Section 5.
%Section \ref{se:highdim}.

Since we wish $\hat \nu_T$ to be small, we therefore require that 
$J(\hat \mf_T,\gamma_T) - \divo{\gamma_T}\sigma_T$ be small. Outside of
$Q$, we have $\divo{\gamma_T}\sigma_T=J(\mf_c,\gamma_c)$. Inside $Q$ 
it is relatively easy
to see that $\sigma_{T}$ has norm and derivatives comparable to 
$\cf_T^{2+q}\sim e^{-nT/2}$. This motivates the 
following conditions on the matter field $\hat \mf_T$.
\begin{definition}
\label{def:mee}
We say that $\hat \mf_T$ satisfies the {\bf momentum error estimates} 
if for each $k$ and 
$\alpha$ there exist constants $C>0$ and $\kappa>\frac{n-1}{2}$ independent 
of $T$ such that
\starteqgroup{M\arabic{equation}}
\begin{equation}
\label{eq:m1}
||J(\hat\mf_T,\gamma_T)-J(\mf_c,\gamma_c)||_{k,\alpha,\Sigma_T\backslash 
\overline{Q}} < C e^{-\kappa T}
\end{equation}
and such that (recalling that $Q_r$ is the centered collar of length $2r$)
\begin{equation}
\label{eq:m2}
||J(\hat\mf_T,\gamma_T)||_{k,\alpha,Q_2} < C e^{-\kappa T}.
\end{equation}
\finisheqgroup\end{definition}
If $\hat\mf_T$ satisfies the momentum error estimates it follows that
\[
||J(\hat\mf_T,\gamma_T) - \divo{\gamma_T}\sigma_T||_{k,\alpha} 
< C e^{-\kappa T}.
\]
The threshold $\kappa>\frac{n-1}{2}$ is the lower limit that will
allow the subsequent analysis of the Lichnerowicz equation to go
forward.  From the momentum error estimates and Lemma \ref{le:vlest}
we immediately obtain
\begin{proposition}
Suppose $\hat\mf_T$ satisfies the momentum error estimates.  Then
there exists a tensor $\hat \nu_T$ on $\Sigma_T$ such that
\[
\divo{\gamma_{T}}(\hat \nu_T) = J(\hat\mf_T,\gamma_T)- \divo{\gamma_{T}} 
(\sigma_T).
\]
In particular, $\hat K_T= \sigma_T+ \hat \nu_T+\frac{\tau}{n}\gamma_T$ 
together with $\gamma_T$ and $\hat\mf_T$ is a CMC solution of the 
momentum and
non-gravitational constraints. Moreover, defining 
$\hat \sigma_T=\hat \nu_T +\sigma_T$, we find that there exist constants 
$C>0$ and $\kappa>\frac{n-1}{2}$ independent of $T$ such that
\begin{equation}
||\hat \sigma_T-\sigma_T||_{k,\alpha} < C e^{-\kappa T}.
\label{eq:sigerr}
\end{equation}
%%% New 10/7/02
\label{prop:sigbound}
%%% End New 10/7/02
\end{proposition}
The method of proof has been outlined above; the details can be found
in \cite{IMP01}.  We note that the existence of $\hat \nu_T$--and therefore 
the existence of $\hat \sigma_T$--follows
even without $\hat\mf_T$ satisfying the momentum error estimates.  
The key point
is that if we impose these estimates, the solution necessarily satisfies 
(\ref{eq:sigerr}).

\subsection{Repairing the Energy Constraint}
\label{se:fixenergy}
Up to this point, we have constructed a CMC solution 
$(\gamma_T,\hat K_T,\hat\mf_T)$
of the momentum and non-gravitational constraints; 
we also have an approximate
solution $\cf_T$ of the Lichnerowicz equation in terms of the 
conformal data $(\gamma_T,\hat K_T,\hat\mf_T)$.  
Our goal is to find a perturbation
$\eta_T$ of $\cf_T$ so that $\cf_T+\eta_T$ solves the Lichnerowicz equation.

Let $\calN_T$ be the Lichnerowicz operator with respect to $\gamma_T$,
$\hat \sigma_T$ and $\hat\mf_T$ on $\Sigma_T$; we write 
\beq
\calN_T(\cf)= \laplaciano{\gamma_T} \cf - a_1 R_{\gamma_T}\cf 
+a_1\left|\hat \sigma_T\right|^2_{\gamma_T}\cf^{-q-3}-a_2\tau^2\cf^{q+1}
+a_1\hat \mnl_T(\cf),
\label{eq:NT}
\eeq
where we abbreviate $n_{\hat \mf_T, \gamma_T}(\psi)$ as $\hat n_T(\psi)$.
Similarly, we let $\calN$ denote the Lichnerowicz operator with 
respect to $\gamma$, $\sigma$ and $\mf$ on $\Sigma$.

Since we will be working with the linearization of $\calN_{T}$,  it is
convenient to denote by $\hat n_{T}'$ the derivative of $\hat n_{T}$ with 
respect to its last argument.  Then the linearization $\calL_{T}$ of
$\calN_{T}$ about $\cf_{T}$ is
\beq
\calL_T= \laplaciano{\gamma_T}  - a_1 R_{\gamma_T} 
+a_1(-q-3)\left|\hat \sigma_T\right|^2_{\gamma_T}\cf_T^{-q-4}-a_2(q+1)\tau^2\cf_T^q
+a_1\ \hat n'_T(\cf_T),
\label{eq:LT}
\eeq
We will similarly denote by $\calL$ the linearization of $\calN$ about 1.

The contraction mapping argument we use here for the existence of a 
solution to the Lichnerowicz equation with data 
$(\gamma_T,\hat K_T, \hat \mf_T)$ requires that the error
$\calE_T:=\calN_{T}(\cf_T)$ decay faster as a function of $T$ than a prescribed
exponential rate; it also requires control of certain mapping properties of
$\calL_{T}$.  The proofs of these
properties follow closely those given in \cite{IMP01},  taking into
account the new features of $\calN_T$ which result from the introduction 
of the term $n_T$, as well as the use of higher dimensions. 
We now outline this argument.

\subsubsection{A Bound for the Error $\calE_T$}
Corresponding to the momentum error estimates of Definition \ref{def:mee},
we make the following definition.
\begin{definition}
\label{def:eee}
Recalling that $\hat n_{T}(\psi_T):=n_{\hat \mf_T, \gamma_T}(\psi_T)$ 
depends implicitly on $\hat \mf_{T}$, and setting 
$n_c(\psi_c):=n_{\mf_c, \gamma_c}(\psi_c)$,  we say that $\hat \mf_T$ 
satisfies the {\bf energy error estimates} if for each $k$ and 
$\alpha$ there exist constants $C>0$ and $\lambda>1/q$ independent 
of $T$ such that
\starteqgroup{E\arabic{equation}}
\begin{equation}
\label{eq:e1}
||\hat n_{T}(\cf_T)-n_{c}(\cf_c)||_{k,\alpha,\Sigma_T\backslash \overline{Q}} 
< C e^{-\lambda T}
\end{equation}
and such that
\begin{equation}
\label{eq:e2}
||\hat n_{T}(\cf_T)||_{k,\alpha,Q_2} <  C e^{-\lambda T}.
\end{equation}
\end{definition}
\finisheqgroup
\begin{proposition}
Suppose $\hat \mf_T$ satisfies the momentum and energy error estimates.  
Then for every $k$ and $\alpha$ there exists a constant $C>0$ and
a constant $\rho>\frac{1}{q}$ not depending on $T$ such that
$||\calE_T||_{k,\alpha} < Ce^{-\rho T}$.
\label{prop:errbound}
\end{proposition}
{\bf Proof:}
The proof follows the approach of the corresponding result in \cite{IMP01}. 
However, the computations are more involved because we wish to prove
the result for $n\ge 3$ and because of
the effects of $\hat \mf_T$ in (\ref{eq:NT}), both directly via $\hat n_T$ and 
indirectly via $\hat \sigma_T$; hence we outline the proof now.

We divide $\Sigma_T$ into three regions, $\Sigma_T\backslash C_T$, 
$C_T\backslash Q$ and $Q$, and estimate the various terms of $\calE_T$ in 
each. For simplicity of presentation, we only discuss estimates of the $C^0$ 
norm of 
$\calE_T$; because of the exponential decay rates of the terms involved, 
the estimates for the higher derivative norms  are essentially the same. 
It is also  convenient to pick in advance the exponent
\[
\rho:=\min\left\{\lambda,\frac{1}{2}+\frac{1}{q},\frac{2}{q},
 2\kappa-1-\frac{3}{q} \right\}.
\]
One can readily verify that $\rho>\frac{1}{q}$.  In fact, the 
restriction $2\kappa-1-\frac{3}{q}>\frac{1}{q}$ is the source of the 
choice $\kappa>\frac{n-1}{2}$ in Definition \ref{def:mee} above.

We first verify the bound in $\Sigma_T\backslash C_T$, where we have 
$\cf_T=\cf_c$, $\gamma_T=\gamma_c$
and $ \sigma_T=\sigma_c$. Since $\cf_c$ solves the Lichnerowicz 
equation with respect to 
$(\gamma_c,K_c,\mf_c)$, it follows that in this region, 
\[
\calE_T=\calN_T(\cf_T) = a_1\left(\left|\hat \sigma_T\right|^2_T-
\left|\sigma_T\right|^2_T\right)\cf_T^{-q-3} + 
a_1\left(\hat n_T(\cf_T)-n_c(\cf_c)\right),
\]
where the definition of $n_c$ is analogous to that of $\hat n_T$.
From Proposition \ref{prop:sigbound} we know that
$\abs{\abs{\hat \sigma_T}^2-\abs{\sigma_T}^2}\le Ce^{-\kappa T}$ in 
$\Sigma_T\backslash C_T$.  From the energy error estimate (\ref{eq:e1}) 
it easily follows that in this region
\begin{eqnarray*}
|\calE_T| &\le& C\max\left(e^{-\kappa T},e^{-\lambda T}\right)\\
&\le& C e^{-\rho T}.
\end{eqnarray*}

Turning now to bounds in the region $C_T$,  we first note the following
easy estimates which hold on all of $C_T$:
\addtocounter{equation}{1}
\starteqgroup{{\arabic{oldeq}\alph{equation}}}
\begin{eqnarray}
\gamma_T & = & ds^2+h+\Ord{e^{-T/2}\cosh(s)} \label{eq:easya}\\
\psi_T & \le & C e^{-T/q}\cosh(2s/q) \label{eq:easyb}\\
\left|\sigma_T\right| &\le & C e^{-nT/2}(\cosh(2s/q))^{q+2} \label{eq:easyc}\\
|\hat \sigma_T-\sigma_T| & \le & C e^{-\kappa T}\label{eq:easyd},
\end{eqnarray}
\finisheqgroup 
where $h$ denotes the round metric on the sphere. In the 
sub-region $Q$, we have (recall that $\psi_T=\psi_1+\psi_2$ in most of $Q_T$)
\[
\calE_T= \calN_T(\cf_T) = (\laplaciano{T}-a_1R_T)\cf_1 + 
(\laplaciano{T}-a_1R_T)\cf_2
+a_1\abs{\hat \sigma_T}^2\cf_T^{-q-3}-q
_2\tau^2\cf_T^{q+1}+a_1 \hat n_T(\cf_T)
\]
From (\ref{eq:easya}) and the definition of $\gamma_T$ it follows 
that $\laplaciano{T}-a_1R_T = \laplaciano{1}-a_1R_1+\Ord{e^{-T/2}}$.  
Hence
\[
\left(\laplaciano{T}-a_1R_T\right)\cf_1 = 
-a_1\abs{\sigma_1}^2\cf_1^{-q-3}+a_2\tau^2\cf_1^{q+1}-a_1n_1(\cf_1)+
\Ord{e^{-\left(\frac{1}{2}+\frac{1}{q}\right)T}},
\]
where $\mnl_1( \psi_1):= \rho (\mConf ({\mf_1}, \psi_1), 
\psi_1^q \gamma_1) \psi_1^{q+1} $.
Since $\mnl_1(\cf_1)=\cf_1^{q+1}\mnl_0(1)$, for 
$\mnl_0(1)=\rho(\mf,\gamma)$, we have the easy bound
$\abs{\mnl_1(\cf_1)}\le C e^{-\left(1+\frac{1}{q}\right)T}$.
Also, since $\abs{\sigma_1}\le C e^{-\frac{n}{2}T}$ and $\cf_j\le C 
e^{-\frac{1}{q}T}$ it follows that
\begin{eqnarray*}
\abs{\left(\Delta_T-a_1R_T\right)\cf_1} &\le& C \max\left( 
e^{-\left(\frac{1}{2}+\frac{1}{q}\right)T}, e^{-\lambda T}\right) \\
& \le & C e^{-\rho T};
\end{eqnarray*}
an analogous estimate holds for $\cf_2$.  

From (\ref{eq:easyc}) and (\ref{eq:easyd}) we determine that
\[
\abs{\hat \sigma_T}^2\le 
C\max\left(e^{-nT},e^{-2\kappa T}\right),
\]
and hence from (\ref{eq:easyb}) we have 
\[
\abs{\hat \sigma_T}^2\cf_T^{-q-3}\le 
C\max\left(e^{-\left(1+\frac{1}{q}\right)T},
e^{-\left(2\kappa-1-\frac{3}{q}\right)T}\right).
\]
From (\ref{eq:easyb}) we also know that 
$\tau^2\cf_T^{q+1}\le C e^{-\left(1+\frac{1}{q}\right)T}$ and from 
(\ref{eq:e2}) that $\abs{\hat n_T(\cf_T)}\le C e^{-\lambda T}$. 
Hence we verify that the estimate
\[
\abs{\calE_T} \le C e^{-\rho T}
\]
holds in the region $Q$.

The remaining region $C_T\backslash Q$ has two components, 
$C_T^{(1)} = [-T/2,1]\times S^{n-1}$ and
$C_T^{(2)} = [1,T/2]\times S^{n-1}$.  By symmetry it suffices to prove
the bound on just one component. In $C_T^{(2)}$ we have 
(recall that $\cf_T=\cf_2+\chi_1\cf_1$ in $C_T^{(2)}$)
\begin{eqnarray}
\calE_T=\calN_T(\cf_T) & = & (\laplaciano{T}-a_1R_T)(\chi_1\cf_1) + 
a_1\left(\abs{\sigma_T}^2\cf_2^{-q-3}-\abs{\hat \sigma_T}^2\cf_T^{q-3}\right) 
\nonumber \\
& & \qquad -a_2\tau^2\left(\cf_2^{q+1}-\cf_T^{q+1}\right)+
a_1\left(\hat n_T(\cf_T)-n_c(\cf_c)\right).
\label{eq:NTC2}
\end{eqnarray}
The last two terms of (\ref{eq:NTC2}) are easy to estimate.  From 
(\ref{eq:e1}) we know
\begin{equation}
\abs{\hat n_T(\cf_T)-n_c(\cf_c)} \le C e^{-\lambda T}.
\label{eq:C2t4}
\end{equation}
Since $\cf_T=\cf_2+\chi_1\cf_1$ and since 
$\chi_1\cf_1=\Ord{e^{-\frac{T}{q}-\frac{2s}{q}}}$ we have
\begin{eqnarray}
\abs{\cf_T^{q+1} - \cf_2^{q+1}} & = &
\Ord{e^{-\left(1+\frac{1}{q}\right)T+\left(2-\frac{2}{q}\right)s}} \nonumber \\
& \le & C \max\left(e^{-\frac{2}{q}T},e^{-\left(1+\frac{1}{q}\right)T}\right).
\label{eq:C2t3}
\end{eqnarray}

We now turn to the second term of the right hand side of (\ref{eq:NTC2}).
From (\ref{eq:easyb}), (\ref{eq:easyc}) and (\ref{eq:easyd}) we have
\begin{equation}
\abs{\abs{\hat \sigma_T}^2-\abs{\sigma_T}^2}\cf_T^{-q-3} \le
C 
\max\left(e^{-\left(\kappa-\frac{1}{q}\right)T},
e^{-\left(2\kappa-1-\frac{3}{q}\right)T}\right).
\label{eq:C2t2p1}
\end{equation}
Also, since
\[
\cf_T^{-q-3} - \cf_2^{-q-3} = 
\Ord{e^{\left(1+\frac{3}{q}\right)T-\left(2+\frac{10}{q}\right)s}}
\]
we have from (\ref{eq:easyc}) that
\begin{equation}
|\sigma_T|^2\left(\cf_T^{-q-3}-\cf_2^{-q-3}\right) \le C 
\max\left(e^{-\frac{2}{q}T},e^{-\left(1+\frac{1}{q}\right)T}\right).
\label{eq:C2t2p2}
\end{equation}
From (\ref{eq:C2t2p1}) and (\ref{eq:C2t2p2}) it then follows that
\beq
\abs{\abs{\hat \sigma_T}^2\cf_T^{-q-3}-\abs{\sigma_T}^2\cf_2^{-q-3}} \le C 
e^{-\rho T}.
\label{eq:C2t2}
\eeq

Hence it remains to estimate 
\[
(\laplaciano{T}-a_1R_T)(\chi_1\cf_1)
.\]
Since $\chi_1\equiv1$ except near $s=T/2$ and since 
$\cf_1=e^{-T/q-2s/q}$ on $C_T^{(2)}$, it follows that
\beq
\chi_1\cf_1 = e^{-T/q-2s/q} + \Ord{e^{-\frac{2}{q}T}}.
\label{eq:C2t1e1}
\eeq
Letting $\gamma_0$ be the round  metric on the cylinder it follows 
from (\ref{eq:easya}) and from calculation of the scalar 
curvature of a round metric that 
\beq
\laplaciano{T}-c_1R_T = \laplaciano{0}-a_1(n-1)(n-2) +\Ord{e^{s-T/2}}.
\label{eq:C2t1e2}
\eeq
Since $\left(\laplaciano{0}-a_1(n-1)(n-2)\right) e^{-T/q-2s/q} =0$,
we see from (\ref{eq:C2t1e1}) and (\ref{eq:C2t1e2}) that
\beq
\left(\laplaciano{T}-a_1R_T\right)\left(\chi_1\cf_1\right) 
 \le C \max\left( e^{-\left(\frac{1}{2}+\frac{1}{q}\right)T},
e^{-\frac{2}{q}T}\right).
\label{eq:C2t1}
\eeq
Hence from (\ref{eq:C2t4}), (\ref{eq:C2t3}), (\ref{eq:C2t2}), and
(\ref{eq:C2t1}) it follows that in $C_T^{(2)}$ (as in the rest of $\Sigma_T$)
\[
|\calE_T|\le C e^{-\rho T}.
\]
\eop

\subsubsection{Mapping properties of the linearization of $\calN_T$}
\label{ssec:mapping}
The goal of this section is to verify certain properties of the
linearization operator $\calL_{T}$: specifically,  that it is invertible 
and that the norm of its inverse on certain weighted spaces is bounded
independent of $T$. These properties, which hold in the vacuum case provided 
$K\not\equiv 0$, depend on the form of the matter term $n(\gamma, \mf, \psi)$ 
and its derivative $n'$ with respect to $\psi$. To guarantee these 
properties, we add the following assumptions to the momentum error 
estimates and the energy error estimates already made:
\starteqgroup{N\arabic{equation}}
\begin{itemize}
    \item For each $p\in\Sigma$,
    \beq
    \label{eq:n1}
    \mnl_{\mf,\gamma}(1,p)-\mnl'_{\mf,\gamma}(1,p)\ge 0.
    \eeq
    \item As $T\rightarrow\infty$,
    \beq
    \label{eq:n2}
    \norm{\hat \mnl'_T(\cf_T)-\mnl'_c(\cf_c)}_{k,\alpha,\Sigma_T\backslash 
\overline{Q}}\rightarrow 0.
    \eeq
    \item As $T\rightarrow\infty$,
    \beq
    \label{eq:n3}
    \norm{\hat \mnl'_T(\cf_T)}_{k,\alpha,Q_2}\rightarrow 0.
    \eeq
\end{itemize}
\finisheqgroup
We use property (\ref{eq:n1})  to establish that $\calL$ (the
linearization of $\calN$ on $\Sigma$ about $\psi=1$) is invertible, and we use 
properties (\ref{eq:n2})  and (\ref{eq:n3}) ensure the convergence of 
$\calL_{T}$ to known operators away from the middle of the 
neck and on the middle of the neck respectively.

To see that $\calL$ is is invertible, we note that
\[
\calL = \laplaciano{\gamma} -a_1\left( R_{\gamma} 
+(q+3)\abs{\sigma}_{\gamma}^2+\frac{a_2}{a_1}(q+1)\tau^2-\mnl'_{\mf,\gamma}(1)
\right).
\]
Since $(\gamma,\sigma,\tau,\mf)$ is a solution of the constraints, we 
can write $R_{\gamma}$ in terms of $\gamma$, $\sigma$, $\tau$ and 
$\mf$ to obtain
\begin{equation}
\calL = \laplaciano{\gamma} -a_1 \left( (q+4) \abs{\sigma}_{\gamma}^2+
q\frac{a_2}{a_1}\tau^2
+\left(\mnl_{\mf,\gamma}(1)-\mnl'_{\mf,\gamma}(1)\right)\right).
\end{equation}
As a consequence of our assumption (\ref{eq:n1}), we know that the term 
in this expression which involves $\mf$ is non-negative.  Moreover,
if the term in parentheses is not identically zero then 
it follows from the maximum principle that $\calL$ has a
trivial kernel on  $\calC^{k+2,\alpha}(\Omega)$.  This is true if 
either $K\not\equiv 0$ or $\mnl_{\mf,\gamma}(1)-\mnl'_{\mf,\gamma}(1)\not\equiv0$;
we will henceforth assume this non-degeneracy condition.
Note that it often holds that $-\mnl'_{\mf,\gamma}(1)\ge 0$.  In this case, 
since $\mnl_{\mf,\gamma}(1)=2\rho(\mf,\gamma)$, the non-degeneracy condition
holds if either $K\not\equiv 0$ or $\rho(\mf,\gamma)\not\equiv 0$.

We now show that for sufficiently large $T$, $\calL_T$ also has 
trivial kernel.  We also want to 
control the norm of its inverse, which we can do in terms of the following 
weighted H\"older space norm: 
\begin{definition}
Let $w_T$ be an everywhere positive smooth function on $\Sigma_T$ 
which equals $e^{-T/q}\cosh(\frac{2s}{q})$ on $C_T$,  which is 
uniformly bounded  away from zero on $\Sigma_R^*$,  and which is equal to
one  on $\Sigma_{2R}^*$. For any $\delta\in \R$, and any $\phi
\in {\mathcal C}^{k,\alpha}(\Sigma_T)$, we set 
\[
||\phi||_{k,\alpha,\delta} = ||w_T^{-\delta}\phi||_{k,\alpha},
\]
and we let ${\mathcal C}^{k,\alpha}_\delta(\Sigma_T)$ denote the
corresponding normed space. 
\label{def:weightnorm}
\end{definition}

\begin{proposition}
Fix any $\delta\in\R$. For $T$ sufficiently large, the mapping 
\[
\calL_T : {\mathcal C}^{k+2,\alpha}_{\delta}(\Sigma_T) 
\longrightarrow  
{\mathcal C}^{k,\alpha}_{\delta}(\Sigma_T) 
\]
is an isomorphism.
\label{prop:nondegenerate}
\end{proposition}
{\bf Proof:}
Since the spaces $\calC_\delta^{k+2,\alpha}(\Sigma_T)$ 
and $\calC^{k+2,\alpha}(\Sigma_T)$ are identical as sets,
and since the associated norms are equivalent to each other
for fixed $T$ and $\delta$, it is enough to show that $\calL_T$ is an
isomorphism for $\delta=0$.  Since $\calL_T$ is Fredholm of index zero
on $\calC^{k+2,\alpha}(\Sigma_T)$, we need only show that, for $T$ 
sufficiently large,  it has trivial kernel.

Suppose not, so that there exists a sequence of increasing parameters $T_k$ 
and  functions $\eta_k$ (not identically zero) such that 
$\calL_{T_k}\eta_k=0$.  We assume without loss of generality 
that $\sup \abs{\eta_k} = 1$.

Suppose now that $\sup\abs{\eta_k}$ is bounded away from zero on 
$\Sigma^*_r$ for some $r\le R$.  As a consequence of  elliptic regularity, 
together with property 
(\ref{eq:n2}), we conclude that there exists a nonzero function $\eta$
on $\Sigma^*$ such that $\calL_c\eta=0$.  Using the conformal 
covariance property 
$\laplaciano{\gamma_c}\eta -a_1R_{\gamma_c}\eta = 
\cf_c^{q+1}\left(\laplaciano{\gamma}(\eta/\cf_c)-a_1R_\gamma\eta/\cf_c\right)$,
we conclude that $\eta/\cf_c$ satisfies
\[
\left[\laplaciano{\gamma}-a_1R_\gamma
-(q+3)a_1\abs{\sigma}^2_\gamma
+a_2(q+1)\tau^2-a_1\cf_c^{-q}\mnl_c'(\cf_c)\right](\eta/\cf_c)=0.
\]
Now, it follows from the definition of $n$ that 
$\mnl_c(\cf)=\cf_c^{q+1}\mnl(\cf\cf_c^{-1})$,
and hence that $\cf_c^{-q}\mnl'_c(\cf_c)=\mnl'(1)$. So $\eta/\cf_c$ solves
$\calL (\eta/\cf_c)=0$ on $\Sigma^*$.  Since $\eta$ is bounded
on $\Sigma^*$ and since $\cf_c$ decays like $r_j^{(n-2)/2}$ near
$p_j$, it follows that $\eta/\cf_c$ is less singular than  
$r_j^{(n-2)}$ at $p_j$ and hence extends to a 
nontrivial solution of $\calL(\eta/\cf_c)=0$ on all of $\Sigma$, 
which is a contradiction.

If on the other hand $\eta_T$ converges uniformly to 0 on $\Sigma^*_r$ 
for any $R>r>0$,
we can consider a sequence of functions on increasing finite length 
sections of the cylinder $\Reals\times S^{n-1}$ by translating $\eta_T$ so
that its maximum occurs at $s=0$.  Then it follows from elliptic regularity 
and the 
properties (\ref{eq:n2}) and (\ref{eq:n3}) that we can extract a subsequence 
that converges in $\calC^2$ on compact sets of the cylinder to a nontrivial 
bounded solution of the equation
\[
\laplaciano{\gamma_0}\eta-\left(\frac{n-2}{2}\right)^2\eta =0.
\]
Since there are no such solutions, we again have a contradiction.
\eop

Let $\calG_T$ denote the inverse of $\calL_T$ on 
$\calC_\delta^{k+2,\alpha}$, which exists for $T$ sufficiently large, 
as a consequence of Proposition 3.  
The proof that $\calG_T$ has bounded norm as $T$ goes to $\infty$
is, using properties (\ref{eq:n2}) and (\ref{eq:n3}), identical to the 
one appearing in \cite{IMP01}, and hence we will not repeat it.  

\begin{proposition}
If  $0 < \delta < 1$, then the norm of the operator
${\calG}_T$ is uniformly bounded as $T \to \infty$.
\label{prop:normbound}
\end{proposition}

\subsubsection{Deforming the conformal factor $\cf_{T}$ to a solution}

We deform $\cfT$ to a true solution $\hat \psi_T=\cfT+\eta_T$ of 
the Lichnerowicz equation using a contraction map.  That is, we
wish to solve
\[
\calN_T(\cfT+\eta_T)=0.
\]
To do this we introduce the quadratically vanishing nonlinearity $\calQ_T$ 
defined by
\[
\calN_T(\cfT + \eta) =\calN_T(\cfT) + \calL_T(\eta) + \calQ_T(\eta).
\]
Then $\cf_T+\eta_T$ is a solution of the Lichnerowicz equation
if and only if $\eta_{T}$ satisfies
\[
\eta_{T} = -\calG_{T}\left(\calE_{T}+\calQ_{T}(\eta_{T})\right), 
\]
and hence is a fixed point of the map $\calT_T$ defined by 
\[
\calT_T(\eta)=-\calG_{T}\left(\calE_{T}+\calQ_{T}(\eta)\right).
\]

The key estimate which leads to a verification  that $\calT_T$ 
is a contraction map 
is that, for $\eta$ near zero, the nonlinear operator $\calN_T$ differs 
from its linear approximation $\calL_T$ by a quadratically small amount.  
That is, 
we wish to show that there is a constant $C$ independent of $T$ such that
\beq
||\calQ_{T}(\eta)||_{k,\alpha,\delta}\leq C ||\eta||_{k,\alpha,\delta}^{2}.
\label{eq:quadest}
\eeq

For the terms in $\calQ_T$ involving $\hat \sigma$ and $\tau$, this 
estimate follows from the explicit form of their dependence on $\eta$. 
To argue this, let us for convenience define 
\[
w_{T}(x) 
:= a_{1}|\hat \sigma_T|^{2}\left(x^{-q-3}-a_2\tau^{2}x^{q+1}\right).
\]
We then have
\[
Q_T(\eta) = w_T(\cf_T+\eta)-w_T(\cf_T)-w_T'(\cf_T)\eta 
+\mnl_T(\cf_T+\eta) -\mnl_T(\cf_T)-\mnl_T'(\cf_T)\eta.
\]
The terms involving $w_T$ can be rewritten in integral form as
\[
w_T(\cf_T+\eta_T)-w_T(\cf_T)-w_T'(\cf_T)\eta_T = 
\int_{0}^{\eta_T}(\eta_T-s)w_{T}''(\cf_{T}+s)\;ds;
\]
hence an estimate of the type (\ref{eq:quadest}) for these terms readily  
follows from appropriate controls on  $w_{T}''(\cf_{T}+s)$.  The only 
serious difficulty stems from the negative exponents appearing in
the definition of $w_T$.  Following \cite{IMP01}, we require that there 
is a constant $c<1$ independent of $T$ such that $|\eta|<c\cf_T$.  This
restriction ensures that $1+\eta/\cf_T$ remains uniformly bounded away 
from zero 
and a straightforward computation using this fact shows that an estimate 
of the form (\ref{eq:quadest}) holds for the $w_T$ terms.

Since there is no a priori form for the dependence of the matter terms 
$n_T$ on $\psi_T$, to complete the estimate  for $Q_T$ we need to make 
a further assumption on $n_T$ (thus extending our list of assumptions 
from section \ref{ssec:mapping}):
\begin{itemize}
\item
\starteqgroup{N\arabic{equation}}
\setcounter{equation}{3}
There exist constants $C>0$ and $c<1$ independent of $T$ such 
that
\beq
\wnorm{\mnl_T(\cf_T+\eta)-\mnl_T(\cf_T)-\mnl'_T(\cf_T)\eta}\le 
C\wnorm{\eta}^2
\label{eq:n4}
\eeq
whenever $|\eta|<c|\cf_T|$.
\finisheqgroup
\end{itemize}

With the estimate (\ref{eq:quadest}) conditionally in hand, we now seek 
an open neighborhood about $0$ on which 
$\calT_T$ (as defined above) is a contraction map.  This neighborhood  
must be mapped to itself under the action of $\calT_T$ and it must be 
small enough so that (for all $T$) it only
contains functions $\eta$ such that $|\eta|<c\,\cf_{T}$ and hence 
the quadratic estimate holds for it.  Fixing $\delta>0$ we find
from Propositions \ref{prop:errbound} and  \ref{prop:normbound} 
that there are constants $C$ and $M$ respectively such that 
$\wnorm{\calG(\calE_{T})}\le C M e^{\left(-\lambda+\frac{\delta}{q}\right)T}$.  
It is therefore convenient to set \[
\Ball_{\nu}=\{u\in\calC^{k,\alpha}_{\delta}
:||u||_{k,\alpha,\delta}\le\nu e^{\left(-\lambda+\frac{\delta}{q}\right)T}\},
\]
and we note for future reference that if $\nu=2CM$ then 
$\calG_T({\calE_{T}})\in B_{\frac{\nu}{2}}$.

Now for $\eta\in\Ball_{\nu}$ we see that inside the neck 
\begin{eqnarray*}
    |\eta| &\le& \nu e^{\left(-\lambda+\frac{\delta}{q}\right) T} 
    \cf_{T}^{\delta-1}\cf_{T} \\
    &\le& \nu e^{\left(-\lambda+\frac{\delta}{q}\right) 
    T}e^{-\frac{T}{q}(\delta-1)}\cf_{T}\\
    &\le& \nu e^{\left(-\lambda+\frac{1}{q}\right)T}\cf_{T};
\end{eqnarray*}
whereas outside the neck
\[
|\eta| \le \nu e^{\left(-\lambda +\frac{1}{q}\right)T}.
\]
Since (we recall from the energy error estimate) 
$\lambda>\frac{1}{q}$, it follows that for any fixed values of 
$c<1$ and $\nu$ there exists
a $T$ large enough so that if $\eta\in\Ball_{\nu}$, then 
$|\eta|<c\,\cf_{T}$ in all $\Sigma_{T}$. Combining this inequality with 
estimate (\ref{eq:quadest}), we find that there 
exists large enough $T$ so that  
\[
\label{quad-diff-est}
||\calQ_{T}(\eta_{1})-\calQ_{T}(\eta_{2})||_{k,\alpha,\delta}\le 
\frac{1}{2M}||\eta_{1}-\eta_{2}||_{k,\alpha,\delta}
\]
for $\eta_{1}$ and $\eta_{2}$ in $\Ball_{\nu}$.  

Setting $\nu=2CM$ we have $\calG(\calE_{T})\in \Ball_{\frac{\nu}{2}}$.
Moreover, setting $\eta_1=\eta$ and $\eta_2=0$ in (\ref{quad-diff-est})
it follows that $\calG(\calQ_T(\eta))\in \Ball_{\frac{\nu}{2}}$ for
$\eta\in \Ball_\nu$ and for sufficiently large $T$.  It follows that
$\calT_T$ maps $\Ball_\nu$ to itself for sufficiently large $T$.  Equation
(\ref{quad-diff-est}) also implies that $\calT_T$ is a contraction
map on $\Ball_\nu$ for $T$ sufficiently large.
(See \cite{IMP01} for further details.) 

If $\calT_T$ is a contraction map, it follows that it has a unique fixed point 
$\eta_T$. It then follows that 
$\hat \psi_T=\psi_T+\eta_T$ satisfies the Lichnerowicz equation 
\beq
\calN_T(\hat \psi_T)=0
\eeq 
with $\calN_T$ defined (for the glued data 
$\gamma_T, \hat \sigma_T, \hat \mf_T, \tau$) as in 
(\ref{eq:NT}). Moreover, the difference between the approximate solution 
$\psi_T$ and the exact one $\hat \psi_T$ is contained in the space 
$\Ball_{\nu}$, which we see from its definition gets arbitrarily small 
as $T$ gets large. We are thus led to our main result: 

\begin{theorem}
Let $(\Sigma^n, \gamma, K, \mf)$ be a smooth, constant mean curvature 
solution of the $n$-dimensional constraint equations for an 
Einstein-matter field theory which satisfies the two criteria of Section 2.  
Let $\Sigma^n$ be compact, and let $p_1$ and $p_2$ be a pair of points in 
$\Sigma^n$.  We assume that this solution data satisfies the following 
conditions: i) The metric is nondegenerate with respect to $p_1$ and 
$p_2$ in the sense of \cite{IMP01}
\footnote{\label{ckftnote}That is that we assume there are no non-trivial 
conformal Killing fields on $\Sigma^n$ which vanish at either $p_1$ or 
$p_2$.}, and either the quantity $K$ or the quantity 
$n_{\mf,\gamma}(1)-n'_{\mf,\gamma}(1)$
appearing in condition (\ref{eq:n1})
is not identically zero.\footnote{Note that for most Einstein-matter field 
theories, it follows from the momentum constraint that if $\mf$ is 
not identically zero, then $K$ is not identically zero.} ii) The momentum
error estimates (Definition 2), the energy error estimates (Definition 3), 
and the matter term estimates (\ref{eq:n1})-(\ref{eq:n4}) all hold. 
Then, for $T$ sufficiently large, there is a one-parameter family of solutions 
$(\Sigma^n_T, \Gamma_T, K_T, F_T)$ of the Einstein-matter field 
constraints with the following properties: The manifolds  $\Sigma^n_T$ 
(all diffeomeorphic to each other) are all constructed by adding a 
handle or neck to $\Sigma^n$,  connecting the two points $p_1$ and 
$p_2$. On the region of $\Sigma^n_T$ outside of the neck, the fields 
$( \Gamma_T, K_T, F_T)$ approach arbitrarily closely to 
$(\Sigma^n, \gamma, K, \mf)$ as $T$ tends to infinity.
\end{theorem}

If we compare the hypotheses of this theorem for gluing solutions of the 
Einstein-matter field constraints with the hypotheses of the corresponding 
gluing theorem for the Einstein vacuum data (See Theorem 1 in  
\cite{IMP01}), it appears as if there are far more conditions which 
must be satisfied  by  the Einstein-matter data. We note, however, that 
for most  of the Einstein-matter theories of physical interest 
(See Section 4), all of the conditions except nondegeneracy are 
satisfied automatically by solutions of the constraints. That is, 
the nature of the matter fields and how they couple into  the 
Einstein-matter constraints for most such theories guarantees 
satisfaction of most of these hypotheses. We shall see this in 
the discussion of example theories in the next section.

\subsection{Gluing in the Presence of a Cosmological Constant}
What happens if we seek to glue solutions 
with a cosmological constant $\Lambda$ 
present? We find that so long as the constant $a_2\tau^2-a_1 \Lambda$ 
is positive (or zero in certain appropriate cases), our results are 
unchanged; otherwise, gluing via the methods we describe here 
generally cannot be carried out. 

To see this, we recall that since the field equations with $\Lambda$ 
present take the form $G=T+\Lambda g$, it follows that the momentum 
constraint (\ref{Jeqn}) is unchanged by the presence of $\Lambda$, 
while the Hamiltonian equation (\ref{Rhoeqn}) is changed to the following:
\begin{equation}
\label{HamLambda }
R_{\gamma}-|K|^2_{\gamma} +(\tr K)^2=2 \rho + 2\Lambda.
\end{equation}
Then, since we require $\Lambda$ to be invariant under the conformal  
action $\Phi$ (otherwise we could not guarantee that it would remain 
a constant) we find that the Lichnerowicz equation with $\Lambda$ 
present takes the form
\begin{equation}
\label{LichneroLambda}
\laplaciano{\lambda} \phi - a_1 R_\gamma\phi +
a_1\left|\sigma + \calD W \right|^2_\gamma \phi^{-q-3} 
-(a_2 \tau^2-a_1\Lambda)\phi^{q+1}
+n_{\mf,\gamma}(\phi)
=0.
\end{equation}
The pairing of $\Lambda$ with $\tau^2$ which we see in (\ref{LichneroLambda}) 
persists throughout the gluing analysis. Since it is crucial to this 
analysis that $a_2 \tau^2-a_1\Lambda\geq0$, our stated result follows. 

\section{Applications to Example Einstein-Matter Field Theories}

The work done in Section 3, culminating in Theorem 1, provides criteria 
for determining whether gluing can be carried out for given solutions of 
the constraints of a specified Einstein-matter field theory. While these 
criteria involve the data of the particular solutions, they depend 
predominantly on the features of the specified theory. In this section, 
we discuss some example field theories for which those solutions of the 
constraints which satisfy the mild nondegeneracy conditions stated in 
hypothesis (i) of Theorem 1 automatically satisfy hypothesis (ii) as well, 
and can therefore be glued.

\subsection{Einstein-Perfect Fluids}

The Einstein-perfect fluid field theories are especially simple from the
 point of view of gluing because they do not require any extra bundle 
structure, they introduce no additional constraints beyond the Hamiltonian 
and the momentum constraints of the vacuum theory, and the fluid 
fields can be readily scaled in such a way that the conformal method 
analysis is without complication. 

The additional fields needed for an Einstein-perfect fluid field theory 
are the fluid energy density $\zeta$, the pressure $p$, and the fluid 
velocity $u$ (which satisfies the condition $g(u,u)=-1$). In terms of 
these quantities, the stress-energy tensor for the field theory is given by 
\beq
\label{Einst-T}
T=\zeta u\otimes u+p(g+u\otimes u);
\eeq
and the field equations consist of \beq
G=T
\eeq
together with
\beq
\div T=0,
\eeq
which is a consequence of  (\ref{Einst-T}). The system is complete, with a 
well-posed Cauchy problem, once an equation of state specifying the pressure
$p$ as a function of the density, $p=F(\zeta)$, has has been given.

Regardless of the choice of equation of state, initial data for a solution 
of the Einstein-perfect fluid equations consists of the vacuum initial data 
$(\Sigma,   \gamma, K)$ together with a scalar field $\zeta$ and a spatial 
vector field $v$ (the spatial projection of the fluid velocity $u$). The 
constraint equations take the form 
\begin{eqnarray}
\label{Jeqn}
\dive K - \extd \tr K = & J &= -(1+|v|^2)^{\frac{1}{2}}
(\zeta+ F(\zeta))v \label{Ein-pf-mom}\\
R_\gamma - |K|^2_\gamma + (\tr K)^2 = & 2\rho& =2 \zeta+ 
2(\zeta + F(\zeta))|v|^2. \label{Ein-pf-Ham}\
\label{Rhoeqn}
\end{eqnarray}

To extend the conformal method to these Einstein-perfect fluid constraints, 
we need to define the conformal action $\Phi$ on the matter fields. While 
one could do this working directly with the fluid initial data fields 
$\zeta$ and $v$, the analysis is much simpler if we work instead with 
the composite functions $\rho$ and $J$. If we are to do this, we need 
the following result:
\begin{lemma} The map from $(\zeta, v)$ to $(\rho, J)$ given by 
$\rho(\zeta, v)= \zeta+ (\zeta +F(\zeta))|v|^2$ and by 
$J(\zeta, v)=-(1+|v|^2)^{\frac{1}{2}}(\zeta+F(\zeta))v$ 
is invertible, so long as we assume the physical conditions 
$ \rho^2\ge |J|^2$,
 $p=F(\zeta)\ge 0$, 
 $F ' (\zeta) <1$, 
and $\zeta\ge 0$.
\end{lemma}
{\bf Proof:}
To prove this lemma, let us presume that we are given $(\rho, J)$, 
and that we want to solve for $(\zeta,v)$. We first notice that it 
follows from  equation (\ref{Jeqn})  that $J$ and $v$ are parallel. 
Hence it is useful to calculate 
$|J|^2=|v|^2(1+|v|^2)(\zeta +F(\zeta))^2=(\rho-\zeta)(\rho+F(\zeta))$, 
and then focus on solving for $(\zeta, |v|^2)$ in terms of 
$(\rho, |J|^2)$.

Let us fix $\rho$ and define the function 
$H_\rho(\zeta):=(\rho-\zeta)(\rho+F(\zeta))$. 
Note from the form of the equation
$\rho(\zeta,v)=\zeta+(\zeta+F(\zeta))|v|^2$ and the hypotheses
that $\zeta\ge0$ and $F(\zeta)\ge0$ be non-negative that if a solution
$\zeta$ exists, we must have $0\le \zeta\le \rho$.
So we seek to solve the equation $H_\rho(\zeta)=|J|^2$ (fixed $J$) 
for $\zeta$ with $0\le \zeta \le \rho$. 
Using the conditions $F(\zeta)\ge0$ 
and $\rho^2\ge |J|^2$ we find
$H_\rho(0)= \rho(\rho+F(0))\ge \rho^2\ge|J|^2$ and also 
$H_\rho(\rho) =0\le |J|^2$. We conclude from the continuity of 
$H_\rho$ that there exists $\zeta \in [0, \rho]$ such that 
$H_\rho(\zeta)=|J|^2$. Calculating then that, as a consequence 
of the condition $F'(\zeta)<1$, we have  $H'_\rho(\zeta)<0$, 
it follows that there is a unique $\zeta(\rho, |J|^2)$ which 
satisfies $H_\rho(\zeta)=|J|^2$. We may use this result together 
with equation (\ref{Rhoeqn}) to solve for $|v|^2 (\rho, |J|^2)$, 
and then finally we may use equation (\ref{Jeqn}) to solve for 
$v({\rho, J})$.
\eop

As a consequence of this Lemma, we may now treat $\rho$ and $J$ as the 
initial data variables for the fluid field, in place of $\zeta$ and $v$. 
Hence, to extend the conformal method to the Einstein-fluid field, we 
define conformal action $\Phi$ on $\rho$ and $J$: we set  
$\Phi( \rho, \phi)=  \rho \phi^{-\frac{3}{2}q-2}$ and 
$\Phi( J, \phi)= J \phi^{-q-2}$.  This choice has three consequences: 
i) The quantity $\frac{\gamma^{ab}J_{a}J_{b}}{\rho^2}$ is conformally 
invariant, and hence the satisfaction of the dominant energy condition 
by the fluid field is also conformally invariant.  ii) The quantity $J$ 
satisfies the condition (\ref{eq:Conf2}), and hence the conformal 
momentum constraint 
\beq
\divo{\gamma}(LW)=\frac{n-1}{n}\phi^{q+2} \nabla \tau +J
\eeq
is independent of $ \phi$ so long as $\tau$ is constant. iii) 
The matter-dependent term in the Lichnerowicz equation
\beq
\laplaciano{\lambda} \phi - a_1 R_\gamma\phi +
a_1\left|\sigma + \calD W \right|^2_\gamma \phi^{-q-3} 
-a_2 \tau^2\phi^{q+1}
+a_1  \rho \phi^{-q/2-1} =0,
\label{Lichneroperfluid}
\eeq
like the $|\sigma+\calD W|$ term, has a positive sign and contains a 
negative power of $\phi$; hence 
its role in the solvability analysis of (\ref{Lichneroperfluid}) is 
essentially the same as the $|\sigma+\calD W|^2_\gamma$ term. Thus we find, 
with this choice of the conformal action on $\rho$ and $J$, the 
conformal method works for the Einstein-perfect fluid constraints 
more or less the same as it works for the Einstein vacuum constraints. 
This is a prerequisite for carrying out the gluing of solutions of the 
constraints as described in Section 3.  

Let us say that we are given a constant mean curvature solution 
$(\Sigma, \gamma, K, \zeta, v)$ of the Einstein-perfect fluid constraints,
together with  pair of points $p_1, p_2 \in \Sigma$ at which we would like 
to carry out a gluing operation. We first rewrite the initial data in the 
$(\Sigma, \gamma, K, \rho, J)$ form; we work entirely in this form until 
the end, at which time we may invert to recover the glued solution in 
$(\Sigma, \gamma, K, \zeta, v)$ form. 

The steps which take us from  $(\Sigma, \gamma, K, \rho, J)$ to the 
preliminary glued data sets $(\gamma_T, K_T, \rho_T, J_T)$ are 
straightforward, as described in Section 3.1.  With no non-gravitational  
constraints $\calC$ present, we set $\hat \rho_T=\rho_T$ and 
$\hat J_T=J_T$, and we next  proceed to repair the momentum constraint, 
as in Section 3.3. To continue with the gluing at this stage, we need 
to check the conformal Killing field (``CK") nondegeneracy condition 
(described in footnote \ref{ckftnote} above) 
and we need to verify the momentum 
error estimates (\ref{eq:m1})-(\ref{eq:m2}). The CK nondegeneracy 
condition is (as with the vacuum case) a mild restriction on the 
class of solutions of the constraints which admit gluing. 
The momentum error estimates, on the other hand, follow immediately 
from the choice of the conformal action $\Phi$ on $(\rho, J)$;
they are automatic for any solution: We verify 

\begin{equation}
||J_T-J)||_{k,\alpha,\Sigma_T\backslash 
\overline{Q}} =0
\end{equation}
and 
\begin{equation}
||J_T||_{k,\alpha,Q_2} < C e^{\frac{n}{2} T}, 
\end{equation}
thereby confirming that these estimates hold.  Consequently the estimate 
(\ref{eq:sigerr}) holds as well. We emphasize  that this is true for any 
chosen solution $(\Sigma, \gamma, K, \rho, J)$ of the Einstein-perfect 
fluid constraints which satisfies the CK nondegeneracy condition relative 
to the chosen gluing points $p_1$ and $p_2$.

We next need to check that the  energy error estimates (\ref{eq:e1}) -
(\ref{eq:e2}) are satisfied. Noting that, for the Einstein-perfect fluid 
theories,  $n_T(\gamma_T, \rho_T, J_T, \psi_T)= \rho_T \psi^{-q/2-1}$,
these estimates are readily verified for any choice of data, much like the 
momentum error estimates just discussed.

To ensure that the linearized Lichnerowicz operator $\calL_T$ is invertible 
for the chosen data $(\Sigma, \gamma, K, \rho, J)$, 
we need the quantity 
$n_{\mf, \gamma}$ to satisfy condition (\ref{eq:n1}). This follows 
immediately from the expression $n_T(\gamma_T, \rho_T, J_T, \psi_T)= 
\rho_T \psi^{-q/2-1\frac{-n}{n-2}}$, regardless of the choice of data. If we 
are working with data on a closed manifold, we also need one or the other 
of the conditions $K \neq 0$ or $\rho \neq 0$ to be satisfied. This is the 
only restriction beyond the CK nondegeneracy condition that we need to make 
on the data itself. Note that it is analogous to the restriction imposed on 
the data for gluing in the vacuum case.      

The remaining assumptions which need to be verified are 
(\ref{eq:n2}), \ref{eq:n3}) and (\ref{eq:n4}). The first two of these 
essentially follow from the conformal invariance of 
$\frac{\gamma^{ab}J_{a}J_{b}}{\rho^2}$. The last of these is a 
straightforward calculation. We thus conclude that any set of 
constant mean curvature initial data satisfying the CK nondegeneracy 
condition relative to $p_1$ and $p_2$ and having either $K$ or $\rho$ 
nonvanishing can be glued at  $p_1$ and $p_2$.

\subsection{Einstein-Maxwell and Einstein-Yang-Mills }
\label{subsec:ym}
We first briefly review the Yang-Mills field theory (on a  fixed 
background spacetime)  to establish our notation. Let 
$\calM$ be a vector bundle with compact structure group $G$ over a 
Lorentz manifold $(M, \eta)$. The sub-bundle of $\mathop{\rm End}(\calM)$ 
consisting of those 
transformations associated with the adjoint representation of $G$
is denoted $\ad(\calM)$; note that each fibre of $\ad(\calM)$ is isomorphic 
to $\lieg$, 
the Lie algebra of $G$.  We fix a bi-invariant metric on $G$,
which induces a metric on the fibres of $\ad(\calM)$. If $\calD$ is 
a connection on $\calM$, then its curvature  $F_{\calD} = \calD ^2$ is a 
2-form 
taking values in $\ad(\calM)$, and therefore is a section of 
$\Omega^2(\ad(\calM))$.

A solution to the Yang-Mills equations is a connection $\cal D$ such
that $\calD^* F_{\cal D}=0$, where $\calD^* = 
(-1)^{\mathop{\rm dim}(M)+1}\hs\cal D\hs$, and
where $\hs$ is the Hodge star operator. (Note that to define this 
operator $\hs$, we need the specified metric $\eta$.)  One readily 
verifies that the Yang-Mills system has a well-posed Cauchy problem, 
for which  the initial data  consists of a $G$ vector bundle $\cal V$ over a 
Riemannian manifold $\Sigma$ with a connection $\dop$ and a section $E$ of 
$\Omega^1(\ad({\cal V}))$.  The variable $E$ plays the part of the time 
derivative of $\dop$ and necessarily satisfies the constraint equation 
$\dAd E = 0$, where $\dAd$ is defined analogously to $\calD^*$ .
In the discussion below, the curvature of $\dop$  appears; 
we denote this curvature by $B_{\dop}$, recalling its familiar role 
in Maxwell's equations. Note that $B_{\dop}$ is a section of 
$\Omega^2(\ad({\cal V}))$

Maxwell's theory is  a special case of Yang-Mills theory, in which  the 
structure group is chosen to be $U(1)$. Since $\lieg$ in this case is 
naturally isomorphic to $\Reals$,  $E$ and $B_{\dop}$ are real-valued 
differential forms. In three dimensions,  $E$ and $\hs B_{\dop}$ are 
the covector fields corresponding to the electric and magnetic fields.

Minimally coupling Yang-Mills to Einstein we obtain the Einstein-Yang-Mills 
theory (with Einstein-Maxwell as a special case), for which the field 
equations are $\calD^* F_{\cal D}=0$ and $G=T$, where $\calD$ is now the 
covariant derivative corresponding to the gravitational as well as the 
Yang-Mills connection, and where the Yang-Mills stress-energy tensor is 
(in component form) 
$T_{\alpha\beta} = F^\mu_\alpha F_{\mu\beta}-\frac{1}{n}g_{\alpha\beta} 
F^{\mu \nu}F_{\mu \nu}$. Using techniques similar to those discussed in 
\cite{EMwellposed},
one readily verifies that the Einstein-Yang-Mills 
field equations have a well-posed Cauchy formulation. A set of initial 
data for Einstein-Yang-Mills consists of a $G$-vector bundle  $\cal V$ 
over an $n$-manifold $\Sigma$, with Einstein data $\gamma$ and $K$, 
together with Yang-Mills data $D$ and $E$, all satisfying the coupled 
constraint equations 
\begin{eqnarray*}
\dive K- \extd \tr K & = & 2(-1)^{n+1}\hs\fip{E}{\hs B_{\dop}} \\
R_\gamma - |K|^2_\gamma + (\tr K)^2 & = & |E|^2+|B_{\dop}|^2\\
\dAd E & = & 0.
\end{eqnarray*}
Here the bilinear map $\fip{\cdot}{\cdot}$ on sections of 
$\Omega^*(\ad(\cal V))$
is the one which is induced from the bi-invariant metric on $\lieg$, 
and the norm on sections
of $\Omega^k(\ad(\cal V))$ is defined by $|W|^2 = \hs <W,\hs W>$.
For Maxwell's equations, the equation $\dAd E = 0$ is the familiar
constraint $\div E = 0$ (the magnetic field constraint $\div B = 0$ is 
absent since it is automatically satisfied when Maxwell theory is 
formulated as a special case of Yang-Mills theory).

The choice of the conformal action $\Phi$ on the Yang-Mills fields $D$ and 
$E$ is essentially determined by the criteria discussed in Section 2. 
In particular, noting that $\gamma_c=\cf^q\gamma$ implies that the
Hodge star operator acting on $k$-forms transforms via
$\hs[c]=\cf^{q\left(\frac{n}{2}-k\right)}\hs$ , we find  that in order 
to satisfy (\ref{eq:Conf1}) and thereby prevent the appearance of $\psi$ 
in the conformal version of the non-gravitational constraint $\dAd E=0$, 
we require that $\Phi(D,\psi)=D$ and $\Phi(E, \psi)= \psi^{-2} E$. It 
then follows that if the conformal data satisfy $\dAd E=0$, then the 
reconstituted data satisfy $\tilde D^* \tilde E=0$. For Maxwell's equations
in three dimensions, the conformal action on the connection 
$\Phi(D,\psi)=D$ is equivalent to a conformal action on the magnetic 
covector field $B$ given by $\Phi(B,\psi)=\psi^{-2}B$.

With this choice of the conformal action, we may proceed to check that 
(\ref{eq:Conf2}) and (\ref{eq:Conf3}) hold as well. We calculate
\begin{eqnarray*}
\hs[c]\fip{\mConf(E,\cf)}{\hs[c]B_{\dop}} & = &
\cf^{q\left(\frac{n}{2}-(n-1)\right)}
\cf^{-2}\cf^{q\left(\frac{n}{2}-2\right)}
\hs\fip{E}{\hs B_{\dop}} \\
&=&\cf^{-q-2}\hs\fip{E}{\hs B_{\dop}},
\end{eqnarray*}
thus satisfying condition (\ref{eq:Conf2}); and we calculate
\[
\cf^{q+1}\left(|\mConf(E,\cf)|_c^2+|B_{\dop}|_c^2\right) = 
\cf^{-3}|E|^2+\cf^{1-q}|B_{\dop}|^2,
\]
from which (\ref{eq:Conf3}) follows, with
\[
\mnl_{E,\dop}(\cf)=\cf^{-3}|E|^2+\cf^{1-q}|B_{\dop}|^2.
\]

To apply the gluing construction, we start with a solution 
$(\gamma,K,E,\dop)$ of the Einstein-Yang-Mills constraints
on a $G$ vector bundle $\cal V$ over $\Sigma$.  From the induced
vector bundle ${\cal V}_c$ over $\Sigma^*$ with connection $\dop[c]$
we construct the vector bundles ${\cal V}_T$ over $\Sigma_T$ as described in 
Section 3 by fixing local trivializations on the 
balls $B_j$ and using them to identify fibres over identified points 
in the connected sum.  We then construct a connection $\dop[T]$ on 
${\cal V}_T$ given by $\dop[c]$ except over $C_T$ where 
$\dop[T]=\chi(s)\dop[1]+(1-\chi(s))\dop[2]$ (here $\chi$ is the cutoff 
function used to define $\gamma_T$).  Hence $\dop[T]=\dop[c]$ 
except over $Q$.
In terms of the classical notation for Maxwell's equations in three dimensions,
the construction of this connection results in 
setting the magnetic field
to $\curl( \chi(s)A_1+(1-\chi(s))A_2)$ over $C_T$, where 
each $A_j$ is a vector potential for the magnetic field over 
the ball $B_j$.

To construct $E_T$, we first note that we may define  local sections 
$E_c$, $E_1$, 
and $E_2$ of the bundle $\ad({\cal V}_T)$ restricted to 
$\Sigma\backslash Q$, $C_T$, and 
$C_T$ respectively by identifying $\ad(\calE_T)$ with $\ad(\calE_c)$ 
over the appropriate subdomains.  We then set $E_T=E_c$ 
over $\Sigma\backslash Q$ and we set
$E_T=\chi_1 E_1+\chi_2 E_2$ over $C_T$,
where $\chi_1 = \chi(t_2-1)$ and $\chi_2=\chi(t_1-1)$.
Thus  $E_T$ is pieced together in the same way that 
the conformal factor $\cf_T$ is (see Section 3.1) and one has 
$E_T = E_1 + E_2$ over most of $C_T$.
It would be more convenient to stitch together $E_T$ over $Q$ alone.
However, doing so results in unacceptably large error terms and would
result in a failure to satisfy the momentum error estimates.

The next step in the gluing construction is the repair  of the 
non-gravitational  constraint: we need to find a section 
$\hat E_T$ of $\Omega^1(\ad(\calV_T))$ which satisfies $\dAd[T]\hat E_T=0$.
To do this, we seek a section
$\mu_T$ of $\ad(\calV_T)$ that satisfies
\[
\dAd[T]\dop[T]\mu_T = -\dAd[T] E_T.
\]
Then $\hat E_T=E_T+\dop[T]\mu_T$ is the section we need.
The operator $\dAd[T]\dop[T]$ is self-adjoint and elliptic; therefore
since $-\dAd[T] E_T$ is $L^2$ orthogonal to the kernel of 
$\dAd[T]\dop[T]$, we can solve this equation.  

To proceed further, we need bounds on $\mu_T$ as well as existence. 
While it is not clear that such bounds always hold, we can prove that 
they do, so long as either the Yang-Mills group $G$ is $U(1)$ (the 
Maxwell case), or so long as $\ad(\calV)$ has no globally parallel 
sections. We note that this last condition holds generically for the 
groups $SU(2)$ and $SU(3)$ of primary physical interest.
The precise statement of the necessary boundedness result is as follows,
where the H\"older spaces used here are defined analogously to those
in Definition \ref{def:holder}. 

\begin{proposition}
Suppose that either $\dop$ acting on $\ad(\calV)$ has trivial kernel 
or $G=U(1)$.  Then the unique solution
$\mu_T$ of $\dAd[T]\dop[T]\mu_T = -\dAd[T] E_T$
satisfies $\norm{\dop[T]\mu_T}_{k,\alpha}<CT^{5/2}\norm{E_T}_{k,\alpha}$ 
for some constant $C$ independent of $T$.
\label{prop:ymbound}
\end{proposition}

We prove Proposition \ref{prop:ymbound} later in this section. For now, 
let us assume the result, and proceed with the verification of  the 
gluing construction conditions. To determine the size of $\mu_T$ and 
therefore the size of the correction $|\hat E_T-E_T|$, we need to 
estimate the quantity $\dAd[T] E_T$ appearing on the right hand side 
of the equation for $\mu_T$. We readily verify that  the following
bounds hold in the regions $Q$ and $C_T^{(2)}$ (Recall from Section 
3.4.1 that $C_T\backslash Q$ can be divided into the two components, 
$C_T^{(1)} = [-T/2,1]\times S^{n-1}$ and
$C_T^{(2)} = [1,T/2]\times S^{n-1}$):
\addtocounter{equation}{1}
\starteqgroup{{\arabic{oldeq}\alph{equation}}}
\begin{eqnarray}
\dop[T] & = & \dop[1]+\Ord{ e^{-\frac{T}{2}+s}}\label{eq:eyma}\\
\hs[T] & = & \hs[1]+\Ord{ e^{-\frac{T}{2}+s}}\label{eq:eymb}\\
E_1 & = & \Ord{e^{-T\left(\frac{n-1}{2}\right)-s(n-1)}}.\label{eq:eymc}
\end{eqnarray}
\finisheqgroup
Now, in the complement of $C_T$ we have $\dAd[T] E_T = 0$, while in  
$Q$, $\dAd[T] E_T = 
\dAd[T] (E_1+E_2)$. Combining (\ref{eq:eyma}), (\ref{eq:eymb}) and 
(\ref{eq:eymc})
together with the identity $\dop[1]\hs[1]\,E_1=0$ we calculate
\begin{eqnarray*}
\dAd[T]E_1 &=& \hs[T]\left(\dop[1]+\Ord{e^{-T/2}}\right)
\left(\hs[1]+\Ord{e^{-T/2}}\right)E_1 \\
& = & \Ord{e^{-nT/2}}.
\end{eqnarray*}
A similar estimate holds for $\dAd[T]E_2$; thence in $Q$, we have 
$\dAd[T]E_T = \Ord{e^{-nT/2}}$.  On $C_T^{(2)}$ we have
$\hs[T]=\hs[2]$, $\dop[T]=\dop[2]$, and $E_T = \chi_1 E_1 + E_2$.
Thus combining (\ref{eq:eyma}), (\ref{eq:eymb}) and (\ref{eq:eymc})
together with the identity $\dop[T]\hs[T]\,E_2=0$ we calculate
\begin{eqnarray*}
\dAd[T] E_T & = & \hs[T]\left(\dop[1]+\Ord{e^{-\frac{T}{2}+s}}\right)
\left(\hs[1]+\Ord{e^{\frac{-T}{2}}}\right)\chi_1 E_1 \\
& = & d\chi_1\ep\left(\hs[1]+\Ord{e^{-\frac{T}{2}}}\right) E_1 + \chi_1 
\hs[T]\left(\dop[1]+\Ord{e^{-\frac{T}{2}+s}}\right)
\left(\hs[1]+\Ord{e^{\frac{-T}{2}}}\right)E_1 \\
& = & \Ord{e^{-T(n-1}} + \Ord{e^{-nT/2}},
\end{eqnarray*}
with an analogous estimate on $C_T^{(1)}$.
So finally, we obtain $\dAd[T] E_T = \Ord{e^{-nT/2}}$, from which 
it follows that  
\[
||\hat E_T-E_T||_{k,\alpha}= \Ord{T^{5/2}e^{-nT/2}}.
\]

Each of the remaining error estimates now needs to be computed.
The techniques are not different from those used above or in the proof of
Proposition \ref{prop:errbound}. Hence we just summarize the results 
of the computation in the following table.
\begin{center}
\extrarowheight=0.2cm
\begin{tabular}{|c|c|c|c|}
\hline
Estimate & $\Sigma\setminus C_T$  & $C_T\setminus Q$ & $Q$ \\ \hline
$\dAd E_T$ & $0$ & $e^{-nT/2}$ & $e^{-nT/2}$ \\ \hline
\ref{eq:m1} \& \ref{eq:m2} & $T^{5/2}e^{-nT/2}$ & $T^{5/2}e^{-nT/2}$ & 
 $T^{5/2}e^{-nT/2}$ \\ \hline
\ref{eq:e1} \& \ref{eq:e2} & $T^{5/2}e^{-nT/2}$ & 
$e^{-T\left(\frac{n-2}{2}\right)}$ & $e^{-T\left(\frac{n+2}{4}\right)}
$ \\ \hline
 \ref{eq:n2} \& \ref{eq:n3} & $T^{5/2}e^{-6T/2}$ & 
 $e^{-T\left(\min\left(1,\frac{n-2}{2}\right)\right)}$ & 
$e^{-T}$ \\ \hline
\end{tabular}
\end{center}
We note in particular that $\hat E_T$ and $D_T$ satisfy the momentum and 
energy error estimates for 
any constants $\kappa$ and $\rho$ for which
$\frac{n-1}{2}<\kappa<\frac{n}{2}$ and 
$\frac{n-2}{4}<\rho<\min\left(\frac{n+2}{4},\frac{n-2}{2}\right)$.

The only remaining gluing conditions which need  to be verified are 
(\ref{eq:n1}) and (\ref{eq:n4}). Since 
\[
n'_{E,\dop}(1) = -4\abs{E}^2+(1-q)\abs{B_{\dop}}^2
\]
it easily follows that for any choice of initial data, 
\[
n_{E,\dop}(1) =  \abs{E}^2+\abs{B_{\dop}}^2\ge n'_{E,\dop}(1),
\]
which establishes (\ref{eq:n1}). As well, it is clear from the form of 
$\mnl'$ that (\ref{eq:n4}) is satisfied for any  $0<c<1$.  Hence we 
conclude that a given set of Einstein-Yang-Mills initial data can be 
glued at a chosen pair of points so long as the CK nondegeneracy 
condition holds for the metric, so long as, if the manifold is compact, 
either $K\neq0, E\neq0$, or $B_D\neq0$, and so long as the hypotheses 
of Proposition \ref{prop:ymbound} are satisfied. 

We now return to the proof of Proposition  \ref{prop:ymbound}. 
We do this first (Lemma \ref{lemma:ymgen}) for the case in which  
$\ad(\Sigma)$ has no parallel sections, and then 
(Lemma \ref{lemma:ymlap}) for the case in which $G=U(1)$.
The key step (at least when no parallel sections are present)
is the establishment of lower bounds 
for the principle eigenvalue of the operator $\dAd[T]\dop[T]$. 

\begin{lemma}
    Suppose  $\ad(\Sigma)$ has no parallel sections.  Then
    there exists a constant $C$ such that for $T$ sufficiently large
    the lowest non-zero eigenvalue $\lambda_T$ of $
    \dAd[T]\dop[T]$ on sections of $\ad(\Sigma_T)$ satisfies
    \[
    \lambda_T\ge \frac{C}{T^2}.
    \]
\label{lemma:ymgen}
\end{lemma}
{\bf Proof:}
If the statement were false, there would exist a sequence of sections
$u_k$ on $\Sigma_{T_k}$ satisfying $\dAd[T_k]\dop[T_k]u_k = \lambda_k 
u_k$ with  $\lambda_k < \frac{1}{kT_k^2}$.  Without loss of 
generality, we can assume $\sup\abs{u_k} = 1$.  Reducing to a 
subsequence and relabeling, we conclude using elliptic regularity 
that $u_k$ converges uniformly on compact subsets of $\Sigma^*$ to a 
solution $u$ of the equation $\dAd[c]\dop[c] u = 0$.

We claim that $u$ is not identically zero.  Suppose that $u_{k}$ converges 
uniformly to $0$ on $\Sigma_T\setminus C_T$. We will show that $u_{k}$
also converges uniformly to 0 on $C_T$, which contradicts the 
assumption $\sup{\abs{u_k}}=1$. First, we note that
\[
\abs{u_k(\theta,t)} = \int_0^t \partial_s \abs{u_k(\theta,s)}\;ds + 
\abs{u_k(\theta,0)}
\]
where the derivatives are meant in the weak sense.  Hence, by H\"older's 
inequality
\[
\abs{u_k(\theta,t)} \le T^{1/2} 
\left(\int_0^T\abs{d\abs{u_k}\,}^2\;ds\right)^{1/2}
+ m_k,
\]
where $m_k:=\sup_{\partial C_T} \abs{u_k}$.
The volume element $dV$ on $C_T$ satisfies the the condition 
$c_1\, dV_0 \le dV \le c_2\, dV_0$ 
for constants $c_1$ and $c_2$ independent of $T$, where $dV_0$
is the volume element of the round cylinder.  Letting $S_t$ be the 
spherical cross-section of $C_T$ at length parameter value $t$,  
we find that 
\begin{eqnarray*}
    \int_{S_t}\abs{u_k}^2
    &\le& C T \int_{C_T} 
    \abs{d\abs{u_k}}^2\;ds + Cm_k^2\\
    & \le& C T \int_{\Sigma_T} \abs{d\abs{u_k}}^2\;dV+Cm_k^2.
\end{eqnarray*}
Now Kato's inequality implies that 
\begin{eqnarray*}
\int_{\Sigma_T} \abs{d\abs{u_k}}^2\;dV 
&\le& \int_{\Sigma_T} \abs{\dop[T_k]u_k}^2\;dV \\
&=& \lambda_k \int_{\Sigma_T} \abs{u_k}^2\;dV, 
\end{eqnarray*}
and hence
\begin{eqnarray*}
    \int_{S_t}\abs{u_k}^2
    &\le& C T \lambda_k \int_{\Sigma_T} \abs{u_k}^2\;dV + Cm_k^2\\
    &\le& C T \lambda_k\Vol(\Sigma_T)+Cm_k^2.
\end{eqnarray*}    
Since $\Vol(\Sigma_T)\le CT$ and since (by hypothesis) 
$\lambda_k\le \frac{1}{kT^2}$,  it follows that
\[
\int_{S_t}\abs{u_k}^2 \le C\left(\frac{1}{k} + m_k^2\right),
\]
and we conclude from interior Schauder estimates applied to 
$\dAd[c]\dop[c]$ that in fact
\[
\sup_{C_T}\abs{u_k} \le C\left( \frac{1}{k} + m_k^2 \right).
\]
Since $m_k$ converges to 0, this proves that $u_k$ converges uniformly
to 0 on $C_T$, which is the desired contradiction.  We conclude that  
$u$ is not identically 0.

We now show that if this non zero limit $u$ exists, 
then it extends to a nontrivial solution of
$\dop u=0$ on $\Sigma$, which contradicts the 
hypothesis that such parallel sections do not exist.
If $\Omega$ is a compact subset of $\Sigma^*$, then
\[
\int_\Omega |\dop u|^2 = \lim_{k\to\infty}\int_\Omega 
\abs{D_{T_k}u_k}^2 \le \liminf_{k\to\infty}\int_{\Sigma_T}
\abs{D_{T_k}u_k}^2 = \liminf_{k\to\infty} \lambda_k\int_{\Sigma_T}
\abs{u_k}^2\le\frac{C}{k},
\]
since $|u_k|\le1$, $\Vol(\Sigma_{T_k})<CT$, and $\lambda_k < 
\frac{1}{kT^2}$.  It follows that $\int_{\Omega} \abs{\dop u}^2 = 0$ 
and we conclude that $\dop u = 0$ on $\Sigma_*$. Since $u$ is bounded, 
it extends to a weak solution of $\dAd\dop u = 0$ on $\Sigma$ and hence $u$ 
is a nontrivial solution of $\dop u=0$ on $\Sigma$, which is a 
contradiction.  We therefore conclude that , for $T$ large enough,  
$\lambda_k\ge \frac{C}{T^2}$.
\eop

The conversion of the eigenvalue estimate of Lemma \ref{lemma:ymgen} 
to the H\"older estimate of Proposition \ref{prop:ymbound}
proceeds in a smilar fashion as was done for the vector Laplacian in 
\cite{IMP01}. From Lemma \ref{lemma:ymgen} we have 
$||\mu_T||_{L^2}\le CT^2||D^*_TE_T||_{L^2}$.  Since the volume of $\Sigma_T$
grows linearly in $T$ we obtain $||D^*_TE_T||_{L^2}\le 
T^{1/2}||D^*_T E_T||_{\calC^{0,\alpha}}$
Finally, local Schauder estimates for $D_{c}^*D_c$ yield 
$||\mu_T||_{C^{2,\alpha}}\le C(||D^*_TE_T||_{C^{0,\alpha}}+||\mu_T||_{L^2})$.
Hence we conclude 
\[
||\dop[T] \mu_T||_{\calC^{1,\alpha}} \le C||\mu_T||_{\calC^{2,\alpha}}\le 
CT^2||D^*_TE_T||_{\calC^{0,\alpha}}\le CT^{5/2}||E_T||_{\calC^{1,\alpha}}.
\]
The estimate of Proposition \ref{prop:ymbound} for higher order spaces 
$\calC^{k,\alpha}$ follows from another application of local Schauder 
estimates.

Lemma \ref{lemma:ymgen} does not apply to $U(1)$ bundles, which do
admit parallel sections.  In this case, $D^*D$ is the Hodge
Laplacian $d^*d$ acting on scalar fields (or equivalently $-\laplacian$ using
our earlier notation and sign convention). The following Lemma implies
Proposition \ref{prop:ymbound} for $U(1)$ bundles.

\begin{lemma}\label{lemma:ymlap}
Any solution $\mu_T$ of $d^*d\mu_T = -d^* E_T$
satisfies $\norm{d\mu_T}_{k,\alpha}<CT^{1/2}\norm{E_T}_{k,\alpha}$ 
for some constant $C$ independent of $T$.
\end{lemma}
{\bf Proof:}
Multiplying both sides of the equation
$d^*d \mu_T = d^* E_T$ by $\mu_T$ and integrating by parts we have
\[
\int_{\Sigma_T} \abs{d\mu_T}^2 \;dV_T = \int_{\Sigma_T} \left<E,d\mu_T\right>\;dV_T.
\]
The Cauchy-Schwartz inequality then implies
\[
||d\mu_T||_{L^2}\le ||E_T||_{L^2}.
\]
Since the volume of $\Sigma_T$ grows linearly we obtain 
\[
||d\mu_T||_{L^2}\le CT^{1/2}||E_T||_{\calC^{0,\alpha}}.
\]
This $L^2$ estimate replaces the eigenvalue estimate of Lemma 
\ref{lemma:ymgen}.  But to use effectively we have to show that $d\mu_T$
satisfies an appropriate elliptic equation.

Applying the Hodge Laplacian $d^*d+d d^*$ to $d\mu_T$ we have
\[
(d^*d+d d^*) d\mu_T  = d d^* d\mu_T = d d^* E.
\]
So from local Schauder estimates applied to the Hodge Laplacian 
we obtain
\begin{eqnarray*}
|| d\mu_T||_{\calC^{2,\alpha}} &\le& 
C( ||dd^*E_T||_{\calC^{0,\alpha}} + ||d\mu_T||_{L^2})\\
&\le & C (||E_T||_{\calC^{2,\alpha}}+T^{1/2}||E_T||_{\calC^{0,\alpha}}) \\
&\le & C T^{1/2} ||E_T||_{\calC^{2,\alpha}}.
\end{eqnarray*}
The proof of the lemma in the case $k=2$ is complete, and the 
the higher order estimates follow from another application
of local Schauder estimates.
\eop

\subsection{Einstein-Vlasov}

The Einstein-Vlasov system is used in general relativity to model
self-gravitating systems of collisionless matter; i.e., matter
which interacts only by means of the collective gravitational field.
There has been a renewed interest recently in establishing rigorous results
for the dynamics  of  solutions of the Einstein-Vlasov system.  Two
useful survey papers on the subject are given in \cite{Ren} and \cite{Andr}.
                                                                
On a Lorentz manifold $(M,g)$ the additional field specified by the
Einstein-Vlasov system is the distribution function representing the
density  of particles with a given space-time position and a given momentum.
Each particle is assumed to travel along a time-like
future directed geodesic, so its momentum at each point is $mv$ where $m$ is
the mass of the particle and $v$ is a future pointing, unit time-like vector.
For simplicity we assume that all the masses are taken to be one; however
the theory can easily be adapted to account for continuous, non-constant,
masses.  The collection 
\[
{\cal P}=\{ (x,v)\, |\, x\in M, v \mbox{  a future pointing, 
unit timelike vector}\} 
\]
forms a Riemannian hypersurface, called the mass shell, 
in the tangent bundle of $M$.  The distribution function is then given by
a non-negative function
\[
f:{\cal P}\rightarrow\R^{+}.
\]
We assume for simplicity that $f$ has compact support 
(again this can be relaxed).
The Einstein-Vlasov system is then
\[
G_{\mu\nu}= T_{\mu\nu}
\]
where we have again chosen units so that the speed of light and $8\pi$
times the gravitational
constant are one. The Einstein-Vlasov stress-energy tensor is given by
\[
T_{\mu\nu}(x) =-\int_{{\cal P}_x} f(x,v) v_{\mu} v_{\nu} dv_{g}
\]
for each $x\in M$, where  $dv_{g}$ is the induced Riemannian volume
measure on the fibre ${\cal P}_x$ of $\cal P$ over $x$.
      
The non-gravitational initial data for the Einstein-Vlasov system 
now consists of a (compactly supported) 
function $f_0$ on the tangent bundle of $\Sigma$.
The energy density and current density of the  non-gravitational  field
are then given by
\begin{eqnarray*}
\rho(x) &=& \int_{T_x\Sigma} f_0(v)\, (1+|v|^2)\, dv_{\gamma}\\
J_{a}(x) &=& \int_{T_x\Sigma} f_0(v)\,\gamma_{ab}v^b dv_{\gamma}
\end{eqnarray*}
where $\gamma$ is the Riemannian metric on $\Sigma$ and $dv_{\gamma}$
is the induced Riemannian volume form on the tangent spaces $T_x\Sigma$.
Note that there are no additional non-gravitational constraints which need 
to be satisfied. 

One readily checks that that if, under the conformal change $\tilde\gamma = 
\psi^q \gamma$, we set
\[
\tilde f_0(v) = \phi^{-\frac32 q -2} f_0(\phi^\frac{q}{2} v)
\]
then
\[
\tilde J_a = \phi ^{-q-2} J_a \qquad \mbox{and} \qquad 
\tilde \rho = \phi^{-\frac32 q -2} \rho.
\]
From this it follows that $n(\phi) = \phi^{q+1}\tilde \rho = 
\phi^{-\frac{q}{2} -1} \rho$ and we observe that the construction 
now proceeds exactly as in the Einstein-perfect fluids case.

\section{Correcting $\sigma_T$ in Higher Dimensions}
\label{se:highdim}

In section \ref{sse:momentum}, the following Lemma was left unproved.

\newcounter{oldlemma}
\setcounter{lemma}{0}
\begin{lemma} Suppose there are no conformal killing fields that 
vanish at points $p_j$ of $\Sigma$.  Then for $T$ sufficiently large and 
for each $X \in {\mathcal C}^{k,\alpha}(\Sigma_T)$ there is a unique
solution $W \in {\mathcal C}^{k+2,\alpha}(\Sigma_T)$ to
$L_{}W = X$.  Moreover, there exists a constant $C$ independent of 
$W$ and $T$ such that
\[
||X||_{k+2,\alpha} \leq C T^{3} ||W||_{k,\alpha}.
\]
\end{lemma}
\setcounter{lemma}{\value{oldlemma}}

The proof of the lemma is identical to the case when $n=3$ found in
\cite{IMP01} so
long as one can establish the following lower bound on the size
of the smallest eigenvalue of the vector Laplacian.

\begin{theorem}
\label{thm:low-eval}
For $T$ sufficiently large, 
the lowest eigenvalue $\lambda_0=\lambda_0(T)$ for $L$ on $\Sigma_T$
satisfies $\lambda_0 \ge CT^{-2}$ for some constant $C$ independent of $T$.
\end{theorem}

In the case $n=3$, Theorem \ref{thm:low-eval} follows from a perturbation 
argument
for the lowest eigenvalue of the vector Laplacian on the round cylinder.
Here again, the proof when $n>3$ is identical to the case when $n=3$ once
one has obtained specific estimates for the vector Laplacian on the
round cylinder, which we turn our attention to now.

For convenience we take the vector Laplacian to operate on 1-forms
rather than on vector fields. Let $X$ be a covector field on the cylinder 
$\Reals\times S^{n-1}$, which we write as
\[
X=f ds + Y(s)
\]
where $Y(s)$ is a covector field on $S^{n-1}$.  The vector Laplacian applied
to $X$ can then be written in terms of its action on  $f$ and $Y$.  We obtain
\[
LX =
L\left(\begin{matrix} f \cr 
Y\end{matrix}\right) =
\left( \begin{matrix} \frac{1-n}{n}\partial_s^2+\frac{1}{2}\Delta_\theta & 
\frac{n-2}{2n}\partial_s\delta_\theta \strut\cr
\frac{2-n}{2n}\partial_sd_\theta & -\frac{1}{2}\partial_s^2 + 
\frac{1}{2}\delta_\theta d_\theta + \frac{n-1}{n}d_\theta \delta_\theta
+ (2-n)\end{matrix}
\right)
\left(\begin{matrix}f \cr Y\end{matrix}\right).
\]

An orthonormal basis of 1-forms on $S^{n-1}$ is given by the eigenfunctions
of the Laplacian.  Moreover, if $\eta$ is a 1-form such that
\beq
\laplacian \eta = \lambda \eta,
\eeq
then from the Hodge decomposition and the topology of the sphere it follows 
that
\beq
\eta = d \phi + \psi
\eeq
where $\phi$ is a function, $\psi$ is a divergence free covector field, and 
both
are eigenfunctions of the Laplacian with eigenvalue $\lambda$.  So if
$\{\phi_j\}$ is an orthonormal basis of eigenfunctions of the scalar Laplacian
with eigenvalues $\lambda_j$, we have an orthonormal basis of 1-forms 
given by $\{\frac{1}{\sqrt\lambda_j}d\phi_j\}\cup\{\psi_j\}$ where 
$\delta\psi_j=0$
and $\laplacian \psi_j=\mu_j \psi_j$.  Finally, we note from \cite{Fo89}
that for each $j$ there exists $k,l\in\N$ such that
$\lambda_j=k(k+n-2)$ and $\mu_j=(l+1)(l+n-3)$. 

For a covector field $X$ of the form 
$u(s)\phi_j ds + v(s)\frac{1}{\sqrt\lambda_j} 
d_\theta \phi_j$ the vector Laplacian acts on the column vector $(u,v)^t$
via
\begin{eqnarray}
L_j' & = & \begin{pmatrix} \frac{1-n}{n} & 0 \cr 0 & -\frac{1}{2} 
\end{pmatrix}\partial_s^2
+\begin{pmatrix} 0 & \frac{n-2}{2n}\sqrt{\lambda_j} \cr -\frac{n-2}{2n}
\sqrt{\lambda_j} & 0 \end{pmatrix} \partial_s + 
\begin{pmatrix} \lambda_j/2 & 0 \cr 0 & \frac{n-1}{n}\lambda_j + (2-n)
\end{pmatrix}
\label{ljprime}
\end{eqnarray}
except when $j=0$.   In this case $\phi_0$ is constant, $X=u(s) ds$,
and we simply have
\[
L_0  =  \frac{1-n}{n}\partial_s^2.
\]
Finally, for covector fields of the form $X=w(s)\psi_j$ we have
\[
L_j''  =  -\frac{1}{2} \partial_s^2 + (\frac{\mu_j}{2} + 2 -n ),
\]
So together
\[
L= L_0\oplus\bigoplus_{j\ge 1} (L'_j \oplus L''_j).
\]

An analysis parallel to that of \cite{IMP01} shows that the temperate solutions
 of $LX=0$ on the cylinder (i.e., those with slower than exponential growth at
both ends) are spanned by $ds$, $s\,ds$, $\omega_{ij}$ and $s\omega_{ij}$,
where for $i<j$, $\omega_{ij}=x_i dx_j - x_j dx_i$ is the generator of
a rotation on the sphere.  So there is a $n(n-1)+2$ dimensional family of
temperate solutions, and the bounded ones are also conformal Killing fields
on the cylinder.  

The remainder of the differences in the analysis are contained in the
following Propositions, which are analogous to Propositions 2 and 3 of 
\cite{IMP01}.

\begin{proposition}
\label{prop:vl-lowev-dirichlet}  Let $\lambda_0=\lambda_0(T)$
denote the first Dirichlet eigenvalue for $L$ on 
$C_T=[-T/2,T/2]\times S^{n-1}$.  Then
\[
\lambda_0(T)\ge \frac{C}{T^2}
\]
for some constant $C$ independent of $T$.
\end{proposition}
{\bf Proof:} The estimate is obvious for $L_0$ and for $L_j''$ when
$\mu_j=2n(n-2)$ (i.e. when $L_j''= -\frac{1}{ 2}\partial_s^2$).  On the other
hand when $\mu_j=(l+1)(l+n-3)$ for $l\ge 2$, the lowest Dirichlet eigenvalue 
of $L_j''$ on $C_T$ converges to $\frac{\mu_j}{2} - (n-2)>0$.  So it remains
to consider the operators $L_j'$.  

Let $X$ denote a lowest eigenfunction 
for $L_j'$ with components $(u,v)$. Then
\begin{eqnarray}
\left<L_j' X,X\right>  & = & \int_{-T/2}^{T/2} \frac{n-1}{n} (u')^2 + 
\frac{1}{2} (v')^2 + 
\frac{n-2}{2n}\sqrt{\lambda_j}(uv'-u'v)+\\&&
\qquad\qquad\qquad\qquad\qquad\qquad
\frac{\lambda_j}{2}u^2+(\frac{n-1}{n}\lambda_j+(2-n))v^2\;ds \nonumber\\
&\ge & \int_{-T/2}^{T/2} \frac{n-1}{n} (u')^2 + \frac{1}{2} (v')^2 + 
\frac{2-n}{n}\left(\frac{(v')^2}{2}+\lambda_j\frac{u^2}{2}\right)+
\nonumber\\&&\qquad\qquad\qquad\qquad\qquad\qquad
\frac{\lambda_j}{2}u^2+(\frac{n-1}{n}\lambda_j+(2-n))v^2\;ds \nonumber\\
&= & \int_{-T/2}^{T/2} \frac{n-1}{n} (u')^2 + \frac{1}{n} (v')^2 + 
\frac{\lambda_j}{n}u^2+(\frac{n-1}{n}\lambda_j+(2-n))v^2\;ds,\label{vl-lowev}
\end{eqnarray}
where we have integrated by parts and applied Young's inequality.
Since $\lambda_j\ge n-1$, it follows that 
$\frac{n-1}{n}\lambda_j+(2-n)\ge\frac{1}{n}$.  Hence,
$\left<L_j' X,X\right>\ge \frac{1}{n}\left<X,X\right> $.
\eop

\begin{proposition} Suppose $LX=\mu X$ on $C_T$.  Then, for 
$T$ sufficiently large, there exists a constant
$c$ independent of $T$  such that if $\mu \le \frac{c}{T^2}$ 
and if
$\int_{S^{n-2}} (\abs{X(-T/2,\theta)}^2+\abs{X(T/2,\theta)}^2)d\theta \le C_1$,
then for any $a\in[-T/2+1,T/2-1]$ we have
\[
\int_{a-1}^{a+1} \int_{S^{n-2}} \abs{X(s,\theta)}^2 \;dsd\theta \le C_2
\]
where $C_2$ depends only on $C_1$ but not $T$ or $a$, and $C_2\rightarrow 0$
as $C_1\rightarrow 0$.
\end{proposition}
{\bf Proof:}
Following the proof of the corresponding result in \cite{IMP01}, it is enough
to prove the result for each component of the separation of variables 
decomposition.  Moreover, the estimate is clear for $L_0$ and $L_j''$,
so we restrict our attention to $L_j'$.  

We can write $L_j'$ in the form
\[
L'_j = - A \del_s^2 + B \del_s + C
\]
where $A$, $B$, and $C$ can be determined from equation (\ref{ljprime}).
Write $X = (u,v)^t$ and $f(s) = \langle AX,X \rangle$.
It follows from the equation $L'_j X = \mu X$ that 
\[
\frac{1}{2}\del_s^2 f(s) = 
|A^{1/2}X' - \frac12 A^{-1/2}BX|^2 + \langle DX ,X \rangle,
\]
where
\[
D = 
\left[ \begin {array}{cc} -1/8\,{\frac { \left( n-2 \right) ^{2}
\lambda_j}{{n}^{2}}}+1/2\,\lambda_j-\mu&0\\\noalign{\medskip}0&-1/16\,{\frac {
 \left( n-2 \right) ^{2}\lambda_j}{ \left( n-1 \right) n}}+{\frac {
 \left( n-1 \right) \lambda_j}{n}}+2-n-\mu\end {array} \right].
\]
Clearly $D_{11}\ge3\lambda_j/8-\mu$.  Also, $D_{22}$ can be rewritten
\[
\frac{(n-1)\lambda_j}{n}\left[1-\frac{1}{16}
\left(\frac{n-2}{n-1}\right)^2\right]+2-n-\mu
\ge \frac{15}{16}\frac{(n-1)\lambda_j}{n}+2-n-\mu.
\]
When $\lambda_j$ is $2n$ (its second non-zero value) we have
\[
\frac{15}{16}\frac{(n-1)\lambda_j}{n}+2-n-\mu = \frac{7n+1}{8}-\mu,
\]
which is positive when $T$ is sufficiently large (forcing $\mu$ 
to be sufficiently small).  Since $D_{22}$ is increasing in $\lambda_j$ we 
have therefore proved that $f(s)$ is convex and hence
the $L^2$ norm of $X$ 
at $s=\pm T/2$ controls the $L^2$ norm of $X$ over any strip $a-1 \leq s 
\leq a+1$ except possibly when $\lambda_j=n-1$. 

To handle the case $\lambda_j=n-1$ we proceed as in Proposition 
\ref{prop:vl-lowev-dirichlet} above. The steps leading to estimate
(\ref{vl-lowev}) follow as before, except now we pick up boundary
terms from the integration by parts.  Specifically, we have
\begin{eqnarray*}
\int_{-T/2}^{T/2}\int_{S^{n-1}}\left<(L_j'-\mu) X,X\right>\;d\theta ds
&=&  \int_{-T/2}^{T/2} \frac{n-1}{n} (u')^2 + \frac{1}{n} (v')^2 + 
\left(\frac{n-1}{n}-\mu\right)u^2\\
&&\qquad\qquad+\left(\frac{1}{n}-\mu\right)v^2\;ds +b(T/2) - b(-T/2)
\end{eqnarray*}
where $b=\frac{1-n}{n}u'u-\frac{1}{2}v'v-\frac{n-2}{2n}\sqrt{n-1}uv$. Hence,
\begin{equation}
\int_{-T/2}^{T/2}\int_{S^{n-1}}
\frac{1}{n}\abs{X'}^2+\left(\frac{1}{n}-\mu\right)\abs{X}^2\;d\theta ds
\le  b(T/2)- b(-T/2).
\label{sig-est-one}
\end{equation}

Let $\chi(s)$ be a cutoff function equal to 0 for $s<0$ and equal to 1 for 
$s>1$, and let $\chi_T(s) = \chi(s-T/2)$.  Then 
\begin{multline*}
\int_{S^{n-1}} \ip<A X'(T/2),X'(T/2)>\;d\theta \quad=\shoveright\\ 
\qquad =\quad
\int_{\frac{T}{2} - 1}^\frac{T}{ 2} \int_{S^{n-1}} \frac{d}{ ds} 
\ip<A X',\chi_T 
X'(T/2)> \;d\theta ds\shoveright\\
\qquad =\quad \int_{\frac{T}{ 2} - 1}^\frac{T}{ 2} \int_{S^{n-1}} 
\ip<A X'',\chi_T 
X'(T/2)> + \ip<A X',\chi_T' X'(T/2)> \;d\theta ds\shoveright\\
\qquad=\quad
\int_{\frac{T}{ 2} - 1}^\frac{T}{ 2} \int_{S^{n-1}} \ip<B X'+ (C-\mu) X,\chi_T 
X'(T/2)> + \ip<A X',\chi_T' 
X'(T/2> \;d\theta ds\shoveright \\
\qquad\le\quad 
\frac{c}{ \epsilon} 
\int_{\frac{T}{ 2} - 1}^\frac{T}{ 2} \int_{S^{n-1}}\abs{X'}^2 + \abs{X}^2
\;d\theta ds 
+ \epsilon  \int_{S^{n-1}} \abs{X'(T/2)} \;d\theta \\
\end{multline*}
where the constant $c$ is independent of $T$.  Taking $\epsilon$ sufficiently
small we obtain 
\begin{equation}
\int_{S^{n-1}} \abs{X'(T/2)}^2 \le c 
\int_{\frac{T}{ 2} - 1}^\frac{T}{ 2} \int_{S^{n-1}} \abs{X'}^2 + 
\abs{X}^2 \;d\theta ds 
\label{sig-est-two}
\end{equation}
where $c$ is independent of $T$. Finally we note that
\begin{equation}
\abs{ b(T/2)} \le \frac{c}{ \epsilon} 
\int_{S^{n-1}} \abs{X(T/2)}^2\;d\theta + \epsilon \int_{S^{n-1}}
\abs{X'(T/2)}^2\;d\theta
\label{sig-est-three}
\end{equation}
for any $\epsilon >0$.  Similar estimates hold at $s=-T/2$, and combining 
(\ref{sig-est-one}), 
(\ref{sig-est-two}), and 
(\ref{sig-est-three}) we conclude there exists a constant $c$ independent
of $T$ such that
\[
\int_{-T/2}^{T/2}\int_{S^{n-1}}
\left(\frac{1}{n}-\mu\right)\abs{X}^2 \le c
\int_{S^{n-1}} \abs{X(-T/2)}^2+\abs{X(T/2)}^2\;d\theta
\]
for $\mu$ sufficiently small, which completes the proof.
\eop

The remainder of the proof of Theorem \ref{thm:low-eval} now 
follows exactly as in \cite{IMP01} and the reader is referred there for 
details.

\section{Conclusions}
Our discussion in section 4 of the Einstein-fluid, Einstein-Yang-Mills,
and Einstein-Vlasov cases provides a sample collection of
Einstein-matter field theories for which solutions of the relevant
constraints can generally be glued together, provided the solutions
satisfy mild nondegeneracy hypotheses. The same is likely true for a
wide collection of other Einstein-matter field theories as well; one
need only check that the conformal method can be applied, and that the
the various criteria discussed in sections 2 and 3 -- and summarized in
Theorem 1 -- are met.
                                                                                
It is not always easy to check these criteria. Indeed, for the
Einstein-Klein-Gordon theory, with the standard spacetime action
principle $S[g,\chi] = \int_M (R +\frac{1}{2} |\nabla \chi|^2 +
\frac{1}{2} m^2 |\chi|^2)$, where $\chi$ is a $\C-$valued scalar field, 
one runs into serious difficulty (even in the massless case when $m=0$).
We note, however, that for this Einstein-matter model, it is also not
straightforward to use the standard conformal method to construct solutions
of the constraint equations.  The difficulty arises in verifying 
solvability for the Einstein-Klein-Gordon version of the Lichnerowicz equation
\beq
\Delta \psi =\frac{1}{8}(R-2|D\chi| ^2) \psi 
-\frac{1}{8}[(\sigma + \calD W)^2+2|P|^2]\psi^{-7} 
 + (\frac{1}{12} \tau^2-
\frac{1}{8}m^2 |\chi|^2) \psi^5,
\eeq
where $P$ is a $\mathbb{C}$-valued 
scalar field, representing  the time derivative of $\chi$.
For solutions of the constraint equations of other field theories, such as the
Einstein-Dirac theory \cite{IOY}, one can readily check whether the
conditions needed for the gluing construction presented here are
satisfied.

For those Einstein-matter theories which do satisfy our gluing criteria,
can we proceed further and obtain stronger gluing results of the sort
established in \cite{CIP-PRL}-\cite{CIP04} (`CIP-gluing') 
for the vacuum case? To be able to show
this, one needs to prove that some appropriate version of the
Corvino-Schoen gluing results \cite{CS03} holds for the field theory of
interest. Corvino appears to have done this for the Einstein-Maxwell
theory \cite{C05}, so we should be able to obtain CIP-gluing
results in this case.


\begin{thebibliography}{9999}


\bibitem{Andr} H.~Andr\'easson,
{\em The Einstein-Vlasov System/Kinetic Theory},
Living Rev. Relativity 5,  (2002), 7. [Online article]:
http://www.livingreviews.org/lrr-2002-7

\bibitem{B88} R. Bartnik,
{\em  Regularity of variational maximal surfaces}, 
Acta Math.  {\bf 161} (1988), 145--181.
          
                                                     
\bibitem{BI04} R. Bartnik and J. Isenberg, 
{\em The Constraint Equations}, in 
The Einstein equations and the large scale behavior
of gravitational fields (P.T.~Chru\'sciel and H.~Friedrich, eds.),
Birkh\"auser, Basel, 2004, 1--39.



\bibitem{Ch70} J. Cheeger {\em A lower bound for the smallest 
eigenvalue of the Laplacian}, in Problems in Analysis, A symposium in 
Honor of S. Bochner, Princeton University Press, 1970, 195--199.    
    
\bibitem{CBY80} Y. Choquet-Bruhat and J. York,
{\em The Cauchy Problem},
in General Relativity and Gravitation - The Einstein Centenary, 
A. Held ed., Plenum 1979, 99--160.


\bibitem{CD02} P. Chru\'sciel and E. Delay,
{\em   Existence of non-trivial, vacuum, asymptotically simple space-times}
Class. Qtm. Grav. {\bf 19}, L71-L79 (2002) 

\bibitem{CD03} P. Chru\'sciel and E. Delay,
{\em On Mapping Properties of the General Relativistic Constraints Operator in 
Weighted Function Spaces, with Applications}
Mem. Soc. Math. de France {\bf 93} (2003), 1--103.

\bibitem{CIP-PRL} P. Chru\'sciel, J. Isenberg, and D. Pollack,
{\em Gluing Initial Data Sets for General Relativity}, 
Phys. Rev. Lett. {\bf 93}, 081101 (2004) 

\bibitem{CIP04} P. Chru\'sciel, J. Isenberg, and D. Pollack,
{\em Initial Data Engineering}, To appear in Comm. Math. Phys.,
gr-qc/0403066 

\bibitem{C00} J. Corvino,
{\em Scalar Curvature Deformation and a Gluing Construction for the Einstein Constraint Equations},
Comm. Math. Phys. {\bf 214} (2000), 137-189. 

\bibitem{C05} J. Corvino, Private Communication

\bibitem{CS03} J. Corvino and R. Schoen,
{\em On the Asymptotics for the Vacuum Constraint Equations},
Preprint (2003), gr-qc/0301071,
To appear J. Diff. Geom.

\bibitem{Fo89} G. B. Folland, {\em Harmonic analysis of the de Rahm 
complex on the sphere}, J. reine angew. Math. {\bf 398} (1989), 130-143

\bibitem{EMwellposed} Y. Four\`es-Bruhat, {\em Th\'eor\'emes d'existence et 
d'unicit\'e pour les équations de la th\'eorie unitaire de Jordan-Thiry},
  C. R. Acad. Sci. Paris  {\bf 232},  (1951). 1800--1802.


\bibitem{I95} J. Isenberg, {\em Constant mean curvature solutions of the Einstein 
constraint equations on closed manifolds}, Class. Qtm. Grav. {\bf 12} (1995), 2249-2274. 
\bibitem{IOY} J. Isenberg,  N. O'Murchadha, and J.W. York, 
{\em Initial-value problem of general relativity III. Coupled fields and the scalar-tensor theory},
Phys. Rev. D {\bf 13}, (1976), 1532-1537. 


\bibitem{IN77} J. Isenberg and J. Nestor, {\em Extension of the York Field 
Decomposition to General Gravitationally Coupled Fields}, Ann. Physics 
{\bf 108} (1977), 368--386

\bibitem{IMP01} J. Isenberg, R. Mazzeo and D. Pollack,
{\em Gluing and Wormholes for the Einstein Constraint Equations},
Comm. Math. Phys {\bf 231} (2002), 529-568.

\bibitem{IMP03} J. Isenberg, R. Mazzeo and D. Pollack,
{\em On the Topology of Vacuum Spacetimes},
Ann. H. Poincar\'e, {\bf 4} (2003), 369-383.

 
\bibitem{Ren} A.D.~Rendall,
{\em The Einstein-Vlasov system},  in 
The Einstein equations and the large scale behavior
of gravitational fields (P.T.~Chru\'sciel and H.~Friedrich, eds.),
Birkh\"auser, Basel, 2004, 231--251.

 
\end{thebibliography}
\end{document}